\documentclass[twocolumn]{emulateapj}

\usepackage{amsmath}
\usepackage{natbib}
\usepackage{graphicx}

\newcommand{\numbersponge}{52}
\newcommand{\numberspongedr}{31}
\newcommand{\numberkok}{9355}

\newcommand{\hi}{H\textsc{i}}

\begin{document}

\title{Recovering interstellar gas properties with \hi\ spectral lines: A comparison between synthetic spectra and 21-SPONGE }

\author{Claire E. Murray\altaffilmark{1$\dagger$}, Sne\v{z}ana Stanimirovi\'c\altaffilmark{1}, Chang-Goo Kim\altaffilmark{2}, Eve C. Ostriker\altaffilmark{2}, Robert R. Lindner\altaffilmark{1}, Carl Heiles\altaffilmark{3}, John M. Dickey\altaffilmark{4}, Brian Babler\altaffilmark{1}}
\altaffiltext{1}{Department of Astronomy, 
                 University of Wisconsin, 
                 Madison, WI 53706, USA}
\altaffiltext{2}{Department of Astrophysical Sciences, 
		Princeton University, Princeton, 
		NJ 08544, USA}
\altaffiltext{3}{Radio Astronomy Lab, UC Berkeley,
		601 Campbell Hall, Berkeley CA 94720, USA}
\altaffiltext{4}{University of Tasmania, 
		School of Maths and Physics, 
		Hobart, TAS 7001, Australia}
\altaffiltext{$\dagger$}{cmurray@astro.wisc.edu}

\begin{abstract}
We analyze synthetic neutral hydrogen (\hi) absorption and emission spectral lines from a high-resolution, three-dimensional hydrodynamical simulation to quantify how well observational methods recover the physical properties of interstellar gas. We present a new method for uniformly decomposing \hi\ spectral lines and estimating the properties of associated gas using the Autonomous Gaussian Decomposition (AGD) algorithm. We find that \hi\ spectral lines recover physical structures in the simulation with excellent completeness at high Galactic latitude, and this completeness declines with decreasing latitude due to strong velocity-blending of spectral lines. The temperature and column density inferred from our decomposition and radiative transfer method agree with the simulated values within a factor of $<2$ for the majority of gas structures. We next compare synthetic spectra with observations from the 21-SPONGE survey at the Karl G. Jansky Very Large Array using AGD. We find more components per line of sight in 21-SPONGE than in synthetic spectra, which reflects insufficient simulated gas scale heights and the limitations of local box simulations. In addition, we find a significant population of low-optical depth, broad absorption components in the synthetic data which are not seen in 21-SPONGE.  This population is not obvious in integrated or per-channel diagnostics, and reflects the benefit of studying velocity-resolved components. The discrepant components correspond to the highest spin temperatures ($1000<T_s<4000\rm\,K$), which are not seen in 21-SPONGE despite sufficient observational sensitivity.  We demonstrate that our analysis method is a powerful tool for diagnosing neutral ISM conditions, and future work is needed to improve observational statistics and implementation of simulated physics.

\end{abstract}

\section{Introduction}

Neutral hydrogen (\hi) in the interstellar medium (ISM) 
plays a crucial role in the life cycles of galaxies.
The atomic medium provides the main fuel reservoir for
molecular gas and, ultimately, star formation. Furthermore, 
the structure of interstellar \hi\ bears important clues to the nature of gas recycling via 
radiative and dynamical feedback and Galactic winds \citep[e.g.,][]{mcg2015}.

Throughout the ISM, \hi\ exists in a ``multi-phase" state, characterized by two thermally-stable phases in pressure
equilibrium \citep{field1969,mckee1977,wolfire2003}: 
the cold neutral medium (CNM) and warm neutral medium (WNM).
An effective constraining observable for the balance between the CNM and WNM
is the excitation temperature (a.k.a., spin temperature, $T_s$) of the gas. However, both
emission and absorption by the 21\,cm hyperfine 
transition of \hi\ are required to measure $T_s$.
Therefore, although the CNM ($T_s\sim 20-200\rm\,K$) has been extensively analyzed 
with \hi\ absorption
\citep[e.g., ][]{crovisier1978, dickey2003, heiles2003},
excellent sensitivity is required to 
constrain the temperature of the WNM ($T_s\sim1000-7000\rm\,K$), 
and few measurements exist
\citep{carilli1998, dwarakanath2002, roy2013, murray2014, murray2015}.

Furthermore, to understand the physical mechanisms responsible for observed \hi\ properties,
comparisons between observations and theory are necessary.
Synthetic datasets from numerical simulations provide a means to 
(1) assess the power of observational diagnostics to reveal the inherent state 
of astronomical systems, and (2) test whether complex simulations 
recover all the properties of real systems.
For example, the velocity structures of synthetic spectral lines provide important
diagnostics of interstellar turbulence \citep[e.g.,][]{falgarone1994}, and the 
nature of CNM dynamics \citep{hennebelle2007, saury2014}.
Furthermore, synthetic observations have been used extensively to
investigate molecule formation \citep{shetty2011, smith2014, duartecabral2015, duartecabral2016}
and Galactic morphology \citep{douglas2010, acreman2012, pettitt2014}.
Important observational biases can be directly quantified using these comparisons.
For example, considering correspondence between the true positions and observed velocities 
of molecular clouds, \citet{beaumont2013} showed that the superposition of clouds along the line of sight introduces
significant uncertainty to observational estimates of cloud mass, size and velocity dispersion.

However, numerical simulations with suitable dynamic range and resolution 
for describing the dynamics of both the CNM and WNM 
have only recently been performed.
\citet{kim2014} constructed a sample of 
synthetic \hi\ absorption and emission spectral lines from 
their three-dimensional hydrodynamical simulations \citep{kim2013}.
Comparing conditions in the simulated data with properties inferred from synthetic spectra,
\citet{kim2014} found excellent agreement between ``true" and ``observed" per-channel and line of sight (LOS)-integrated properties such as column density and spin temperature.
Furthermore, they showed that column densities computed in the optically-thin limit
significantly underestimate the true column density when the \hi\ optical
depth is greater than $\tau \sim 1$. This agrees with previous comparisons of observed
and simulated LOS column density by \citet{chengalur2013}, who used Monte Carlo simulated spectra
to test the role of optically-thick \hi. However, \citet{kim2014} found the discrepancy factor to be much smaller than  
\citet{chengalur2013}, indicating that when proper dynamics are considered, spectral line blending due to overlapping cold clouds 
is not significant. 

\begin{figure}
	\vspace{-20pt}
   \includegraphics[width=0.5\textwidth]{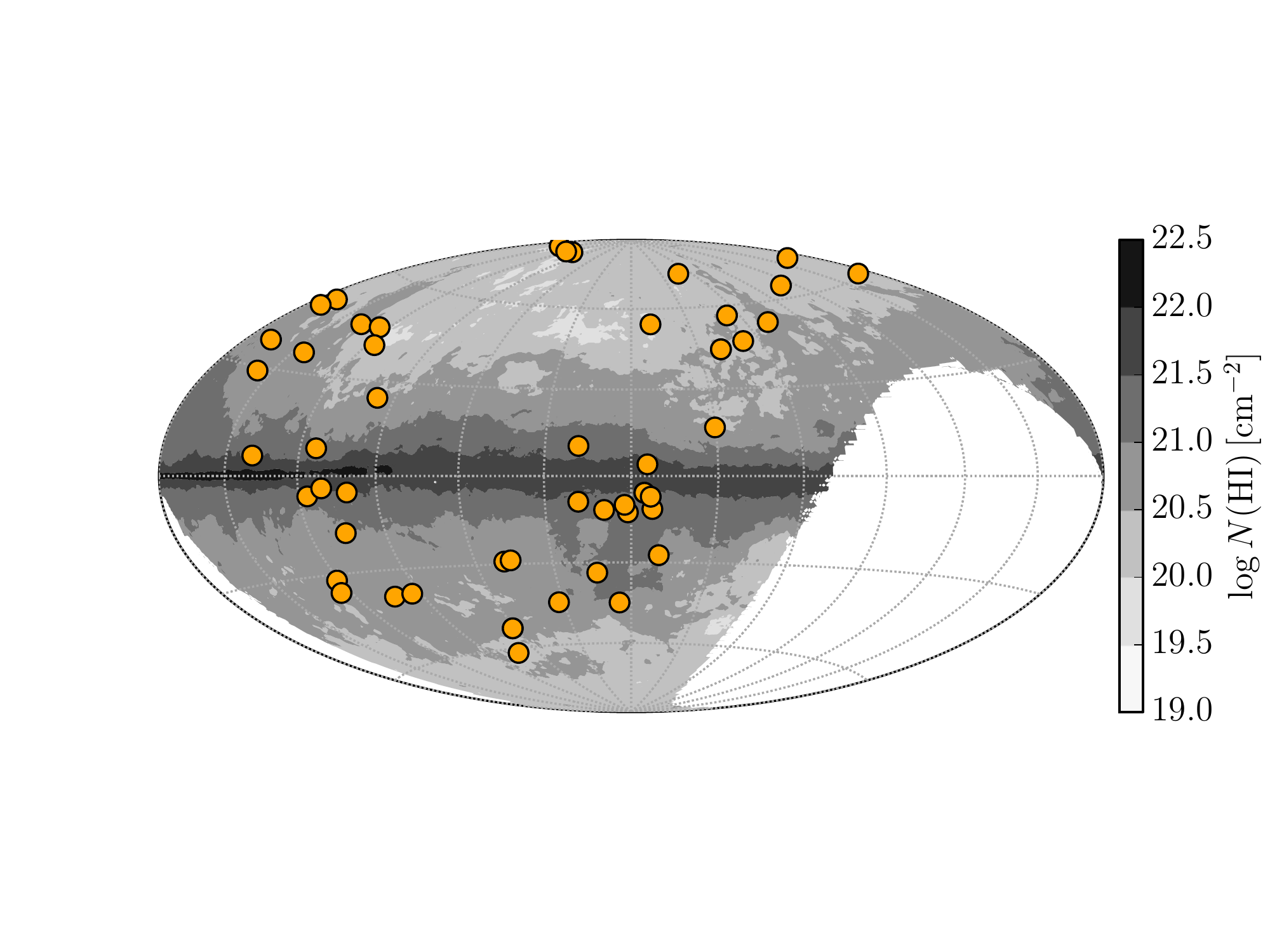}
        \vspace{-40pt}
   \caption{Galactic \hi\ column density ($N({\rm H\textsc{i}})$) map from the Leiden Dwingeloo Survey \citep{hartmann1997}. The locations of the \numbersponge{} 21-SPONGE \hi\ absorption line sources (orange circles) are overlaid.}
   \vspace{10pt}
      \label{f:map}
\end{figure}

Although LOS-integrated ISM properties provide important diagnostics, 
interpretation is more complicated for multi-temperature, as opposed to isothermal, conditions.
Observations show that even high galactic latitude lines of sight contain
several components of varying spin temperature \citep[e.g.,][]{heiles2003, begum2010, roy2013, murray2015}.
The technique of Gaussian decomposition is one method that can identify
individual spectral components from disparate \hi\ phases,
and has been used extensively to disentangle complex 
spectral lines \citep[e.g.,][]{lazareff1975,mebold1982,dickey1990,mohan2004}.
When applied to both \hi\ absorption and emission spectra, Gaussian decomposition can be used
to estimate the spin temperatures of individual spectral features and the 
fraction of CNM along the line of sight \citep{heiles2003,stanimirovic2014, murray2015}.
However, the method suffers from non-uniqueness complications, as Gaussian functions 
do not form an orthogonal basis \cite[e.g.,][]{heiles2003}
and the number of Gaussian functions 
used to produce reasonable spectral fits can vary significantly.
For example, comparing the Gaussian decompositions of 
the Galactic \hi\ absorption spectrum towards 3C138, 
there are no fitted components which agree between 
the 6 found by \citet{murray2015} and the 13 
found by \citet{roy2013}.
Furthermore, no quantitative estimates of how well
Gaussian functions recover the properties of interstellar gas exist.

One of the major scientific goals for future \hi\ observations with 
the Australian Square Kilometer Array Pathfinder (ASKAP) and the
Square Kilometer Array (SKA) will be to understand the temperature 
distribution of the ISM and how it relates to the life cycles of galaxies 
using thousands of spectra \citep[e.g.,][]{dickey2013, mcg2015}. 
However, the first step in this undertaking is to understand the
biases and limitations of our observational and analysis methods in
reproducing interstellar gas properties. This is best done
by analyzing synthetic \hi\ data from numerical simulations,
which include realistic physical processes, and also provide 
full 3D information on simulated \hi\ structures (e.g. density, temperature, velocity).

Accordingly, we begin this paper by quantifying how well Gaussian analysis
of \hi\ spectral lines via simple radiative transfer
recovers ``true'' interstellar gas properties by analyzing 
synthetic $21\rm\, cm$ spectral profiles derived from 3D hydrodynamical simulations from \citet{kim2013,kim2014}. 
To analyze the synthetic spectral profiles, we use the Autonomous Gaussian Decomposition (AGD) algorithm \citep{lindner2015}.
AGD implements derivative-based computer vision to perform Gaussian decomposition of 
spectral lines, enabling efficient, reproducible and objective spectral decomposition

In the second half of the paper, we compare synthetic
and observed \hi\ spectra objectively using the same methodology. 
For this we use data from the 21-cm Spectral Line Observations of Neutral Gas with 
the VLA (21-SPONGE) survey, one of the most sensitive
Galactic $21\rm \, cm$ surveys \citep{murray2015}, as well as the Kim et
al. synthetic \hi\ spectra. 
We assess the ways in which the detailed statistical properties 
of synthetic spectra may agree or disagree with the statistics of observed spectra.  
This in turn reflects the influence of the star formation feedback 
mechanisms and other physics of the simulations.
We especially focus on the importance of Ly$\alpha$ resonant scattering for
\hi\ excitation and the temperature distribution.

In Section~\ref{s:obs}, we describe the 21-SPONGE observations, and
in Section~\ref{s:sims} we describe 
Kim et al. simulations and synthetic data.
We present and discuss our analysis method in Section~\ref{s:agd}.
In Section~\ref{s:sim_an}, we compare the properties inferred from synthetic spectra with 
the simulated properties of gas along the same LOS. We then compare the
synthetic spectra with 21-SPONGE observations in Section~\ref{s:agd_compare}.
Finally, we present our summary and conclusions in Section~\ref{s:conclusions}.

\section{Observations}
\label{s:obs}

For observations of Galactic \hi, we use data from the 21-SPONGE survey \citep{murray2015}. 
21-SPONGE is the most sensitive survey of Galactic \hi\ absorption at the Karl G.
Jansky Very Large Array (VLA) to date. We target strong extragalactic
radio continuum sources mostly at high Galactic latitudes ($|b|>5^{\circ}$), and consistently reach RMS noise
in \hi\ optical depth of $\sigma_{\tau} < 10^{-3}$ per $0.4\rm\,km\,s^{-1}$ channels. 

21-SPONGE utilizes \hi\ emission observations along the same LOS 
from the Arecibo Observatory ($\sim3.5'$ beam at $21\rm\,cm$). 
The emission spectrum in the direction of each source is computed by observing a pattern of 16
off-source positions and interpolating across the target position 
\citep[see, e.g., ][hereafter HT03]{heiles2003}. 
We note that the Arecibo \hi\ emission profiles are not corrected for an 
effect known as ``stray radiation", wherein radiation enters the main telescope beam 
through higher-order side lobes. 
Although stray radiation can be modeled and removed from \hi\ emission data
\citep[e.g., LAB and GASS surveys; ][]{kalberla2005, mcg2009, kalberla2010}, it is a complex 
process requiring stable beam shapes which are not achieved at Arecibo.
Comparison between the GALFA-\hi\ survey at Arecibo and the stray radiation-corrected LAB survey 
suggested that stray radiation likely does not contribute more than $\sim500\rm\,mK$ over $\sim50\rm\,km\,s^{-1}$ to
observed \hi\ brightness temperature \citep{peek2011}. We emphasize that the effect is only significant for emission, not absorption.

Armed with high-sensitivity \hi\ absorption and emission spectra along each line
of sight, 21-SPONGE is sensitive to \hi\ column densities and 
temperatures from all neutral ISM phases, 
including the CNM, WNM and thermally unstable medium. 

In \citet{murray2015}, we presented the survey design, analysis methods
and initial results for 21-SPONGE. To derive physical properties of 
interstellar gas along each line of sight, we decomposed (by hand) \hi\ absorption and emission 
spectral pairs simultaneously into Gaussian functions, 
and solved radiative transfer equations to derive
the column density and spin temperature of individual spectral components, taking into account the
presence of self-absorption and the order of features along 
each line of sight \citep[as done, e.g., in HT03, ][]{stanimirovic2005,stanimirovic2014}.
We found excellent agreement with previous \hi\ absorption surveys.
The high sensitivity of 21-SPONGE allowed us to extend the
maximum \hi\ spin temperatures detected directly in absorption and emission
from $600\rm\,K$ in HT03 to $\sim1500\rm\,K$ \citep{murray2015}. 

Figure~\ref{f:map} displays the positions of the \numbersponge{} 21-SPONGE sources
overlaid on an \hi\ column density map from the Leiden Dwingeloo Survey \citep[LDS; ][]{hartmann1997}.
The targets probe a large range in Galactic latitude.

\section{Numerical simulations}
\label{s:sims}

We analyze recent high-resolution Galactic ISM simulations 
by \citet[hereafter KOK13]{kim2013}. These simulations include momentum feedback 
from supernovae, time-varying heating, interstellar cooling appropriate for warm/cold gas, 
galactic differential rotation, 
gaseous self-gravity and external gravity from dark matter and stars. We refer the reader to 
KOK13 for a full description of the numerical setup, and methods.

Using these simulations, \citet[hereafter KOK14]{kim2014} constructed a 
set of synthetic spectral lines sampling the
local ISM. Assuming an observer sits in the center of the simulation,
they selected $10^4$ positions randomly distributed in Galactic
latitude ($l$) and longitude ($b$) and extracted the number density ($n$), 
temperature (kinetic, $T_k$, and spin, $T_s$) and velocity ($v$) as functions of path length ($s$).
These lines of sight (LOS) are restricted to $|b|>4.9^{\circ}$ and
$s\leq3\rm\,kpc$ 
so that the limited horizontal extent of the simulation does not adversely affect the results, as would be a concern at
low Galactic latitude. 
For the particular KOK13 simulation we analyze in this work (the model denoted ``QA10"), 
galactic rotation was applied with assumed angular velocity of $\Omega = 28$ km sec$^{-1}$
and gas surface density $\Sigma= 10$ M$_{\odot}$ pc$^{-2}$ (KOK13). 

For each observed LOS, KOK14 produced synthetic \hi\ $21\rm\,cm$ emission and absorption
as functions of radial velocity using analytical radiative transfer and a simple prescription for line excitation
prescriptions. The reader is referred to Section 2.3 of KOK14 for a complete description 
of the methods used to construct the synthetic spectra. 

\begin{table*}
\caption{AGD Summary}
\centering
\begin{tabular}{lc|cc|cc||cc}
\hline
\hline
Source            & LOS   &  \multicolumn{2}{c|}{Absorption ($N_{\rm AGD}$)} & \multicolumn{2}{c|}{Emission ($N_{\rm AGD}$)}  & \multicolumn{2}{c}{Matched$^{\rm b}$}   \\
                       &  (number)                       &  (total)  & (per LOS)$^{\rm a}$    &  (total) &  (per LOS)$^{\rm a}$    & (total) &  (per LOS)$^{\rm a}$    \\
\hline
\hline
KOK14                &  \numberkok{}          &   14023        & $1.5 \pm 1.7$  &    23475     &  $2.5 \pm 1.4$   &  9218        &   $1.0 \pm 1.1$  \\
KOK14  (no WF) &  \numberkok{}          &   15468        & $1.7 \pm 1.8$  &    23519     &  $2.5 \pm 1.4$   &  9490        &   $1.0 \pm 1.1$  \\
\hline
21-SPONGE   &  \numbersponge{}        &    237 	   & $4.6 \pm 3.0$  &    326         &  $6.3 \pm 2.9$   &  88             &   $1.7 \pm 1.3$  \\

\hline
\label{t:AGD}
\end{tabular} \\
\footnotesize \raggedright $^{\rm a}$: Mean and standard deviation over all LOS  \\
\footnotesize \raggedright $^{\rm b}$: ``Matched" statistics will be discussed in Section 6.
\end{table*}

In particular, as part of their model for synthetic $21\rm\,cm$ level populations, KOK14
considered indirect radiative transitions due to resonant scattering by Ly$\alpha$ photons
\citep[the Wouthuysen-Field (WF) effect; ][]{wouthuysen1952, field1959}
in addition to collisions and direct radiative transitions.
They parameterized the WF effect following \citet{field1959} with a constant value for the 
Galactic Ly$\alpha$ photon number density, $n_{\alpha}$, inferred from \citet{liszt2001} to be $n_{\alpha}=10^{-6}\rm\,cm^{-3}$.
This value of $n_{\alpha}$ is highly uncertain and difficult to constrain observationally or numerically.
Given that observed LOS-integrated and per-channel properties are dominated
by high-optical depth gas, wherein the $21\rm\,cm$ transition is already thermalized by collisions due to high densities,
the WF effect does not significantly affect these values. 
The WF effect should be most important for the WNM, where generally $T_s\leq T_k$ due to 
the inefficiency of collisions at thermalizing the $21\rm\,cm$ transition \citep[e.g., ][]{liszt2001}. 
Indeed, at high $T_s$ the WF effect is significant (c.f., Figures 9a, 10a of KOK14).

We note that the KOK13 simulations do not include chemistry and the \hi-to-H$_2$ conversion. In addition, their
implemented supernova feedback injects momentum and not thermal energy, resulting in the
absence of a hot ($T\sim10^5-10^7\rm\,K$) medium. Although for the warm and cold medium these are secondary effects, 
and they are being addressed in ongoing simulations with thermal supernova feedback to create a hot ISM
\citep{kim2016sub}, we need to keep these limitations in mind when considering properties of synthetic spectra from KOK14.

In this paper, we analyze components within the synthetic KOK14 \hi\ spectral pairs
to investigate how radiative transfer-based Gaussian fitting reproduces real physical quantities. 
From their catalog of $10^4$ spectral pairs, we selected those without saturated (defined here as $\tau \geq 3$) or NaN-valued absorption lines, for a final catalog of \numberkok{} \hi\ spectral pairs. To simulate observational conditions, we added Gaussian-distributed noise with an amplitude
of $\sigma_{\tau}=10^{-3}$ to each absorption spectrum (equal to the median
RMS noise in $\tau$ per channel from 21-SPONGE) and 
$\sigma_{T_{\rm B}}=0.2\,\rm K$ to each emission spectrum (equal to the median
RMS noise in $T_{\rm B}$ per channel in 21-SPONGE). 

\section{Gaussian Decomposition with AGD}
\label{s:agd}

To perform Gaussian fits to \hi\ spectra (either real or synthetic),
we use the Autonomous Gaussian Decomposition algorithm
\citep[AGD; ][]{lindner2015}.
AGD implements derivative spectroscopy and machine learning techniques
to efficiently and objectively provide initial guesses (i.e., amplitude, width, mean velocity)
for multiple-Gaussian fits to spectral line data. 

Before implementing AGD, we trained the algorithm to maximize the decomposition accuracy.
We began by constructing a synthetic \hi\ dataset from the Gaussian components detected by the 
Millennium Arecibo 21\,cm Absorption Line Survey \citep[HT03, ][]{heiles2003b}.
The synthetic training dataset construction and training
are described fully in Lindner et al.\,(2015), and summarized here for clarity.
We selected the number of components in each synthetic spectrum to
be a uniform random integer ranging from the mean 
number of components in the survey (3) to the maximum number (8; HT03), 
and then drew the component parameters from the published HT03 amplitude, 
FWHM and mean velocity distributions with replacement.
As done with the KOK14 synthetic spectra, we added Gaussian-distributed noise with an amplitude
of $\sigma_{\tau}=10^{-3}$ to each absorption spectrum (equal to the median
RMS noise in $\tau$ per channel from 21-SPONGE) and 
$\sigma_{T_{\rm B}}=0.2\,\rm K$ to each emission spectrum (equal to the median
RMS noise in $T_{\rm B}$ per channel in 21-SPONGE). The synthetic training
sets for absorption and emission consist of 20 spectra each.

After constructing the synthetic training dataset, we used the Python implementation of 
AGD, GaussPy, 
 to decompose the synthetic training dataset
for different values of the ``two-phase" smoothing parameters $\alpha_1$ and $\alpha_2$. 
These smoothing parameters serve to identify the types of spectral properties present in the data. 
Beginning with initial choices for $\alpha_1$ and $\alpha_2$ and 
a signal-to-noise (S/N) threshold below which the algorithm will not select 
components (S/N=3.0), GaussPy computes the accuracy of the decomposition 
(i.e. how closely the derived model parameters are to the true model parameters), 
for iteratively different values of $\alpha_1$ and $\alpha_2$ until it converges on 
minimal model residuals and maximum decomposition accuracy. 
After training, we found accuracies of
80\% and 70\% for \hi\ absorption and emission decompositions, respectively. The resulting 
values are $\alpha_1=1.12$ and $\alpha_2=2.73$ for absorption and  $\alpha_1=1.70$ and $\alpha_2=3.75$
for emission.

With trained values of $\alpha_1$ and $\alpha_2$ in hand, we used GaussPy to apply the AGD algorithm identically 
(i.e. same values of $\alpha_1$, $\alpha_2$ and S/N) to the 
observed 21-SPONGE and simulated KOK14 spectra to 
derive lists of Gaussian parameters for each dataset. Table~\ref{t:AGD}
summarizes the decomposition results for the emission and absorption 
spectra from 21-SPONGE and KOK14 including
the WF effect and without the WF effect (``no WF").
The typical uncertainties in the fitted parameters are 
$\sim1-10\%$ from the least-squares fit applied by AGD.

\section{Assessing the power of the Gaussian-fitting method with synthetic \hi\ spectra}
\label{s:sim_an}

From the KOK14 simulations, we have information about the density and spin temperature as a function of distance
($n(s)$, $T_s(s)$), as well as the optical depth and brightness temperature as a function of velocity ($\tau(v)$, $T_B(v)$)  
for each line of sight. Therefore, following AGD analysis, we can 
compare inferred properties from spectral lines to the true gas properties within the simulated ISM. This will allow
us to quantify the biases and limitations of Gaussian analysis in reproducing 
realistic physical properties.

\begin{figure*}
   \includegraphics[width=\textwidth]{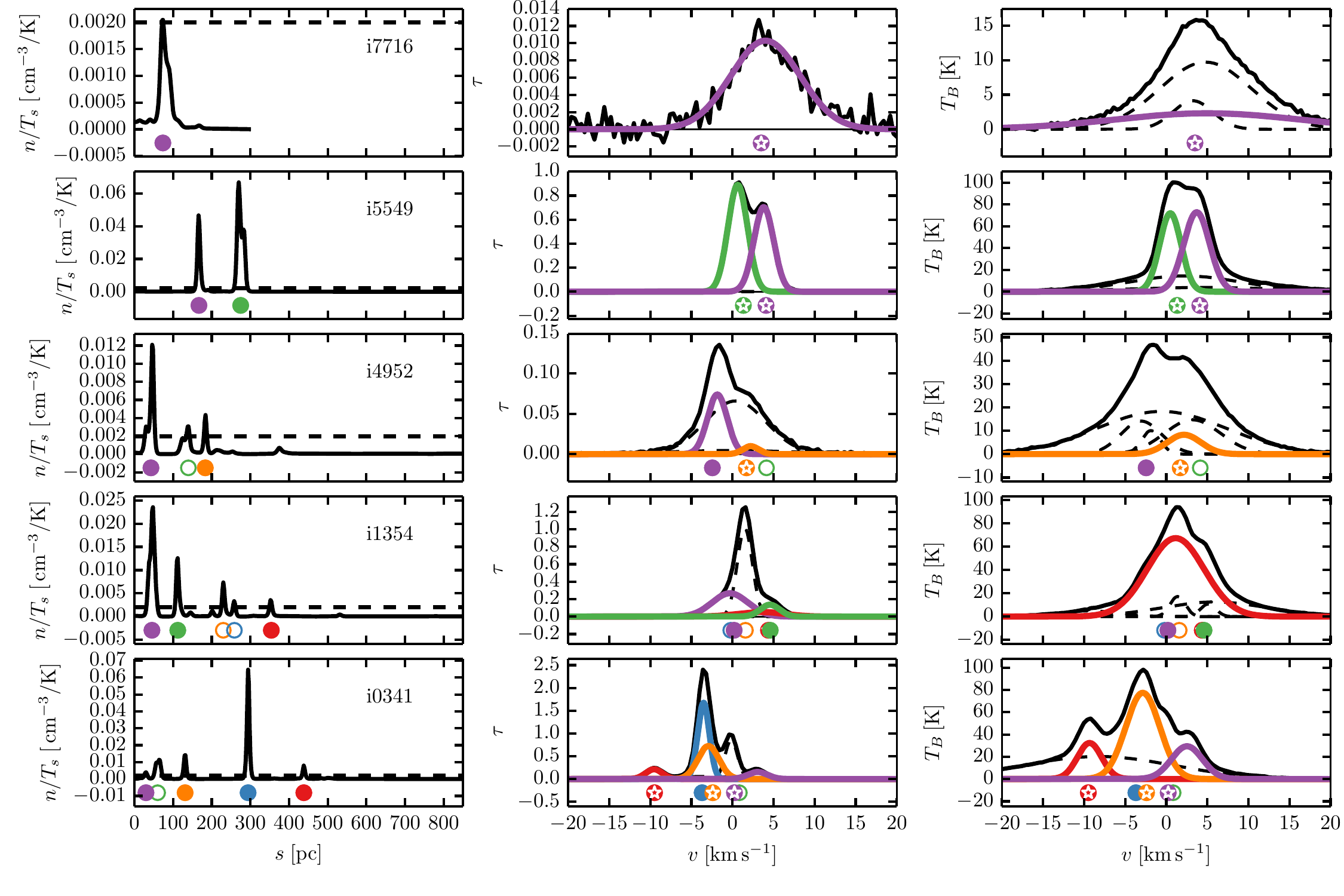}
   \caption{Example LOS from KOK14. Left: density over spin temperature ($n/T_s$) as a function of distance (s) along the LOS; Middle: synthetic optical depth as a function of velocity ($\tau(v)$); Right: synthetic brightness temperature as a function of velocity ($T_B(v)$). Gas structures defined by peaks above $(n/T_s)_{\rm thresh} = 0.002\rm\,cm^{-3}/K$ (dashed line, left column) are indicated by colored circles in each panel. If a structure from the left column matches with an AGD absorption component in the middle column according to Equations~\ref{e-sigs2} and~\ref{e-fwhm2}, the circles are filled and the matching AGD component is plotted in the corresponding color (unmatched components are plotted in dashed lines). If a structure matches with an AGD absorption \emph{and} emission component according to Equations~\ref{sigs_away} and~\ref{fwhm_factor}, the circle symbol has a white star within it.}
   \vspace{10pt}
      \label{f-cloud-matched-ex}
\end{figure*}

\subsection{Defining gas structures in position and velocity}

\begin{figure}
   \includegraphics[width=0.5\textwidth]{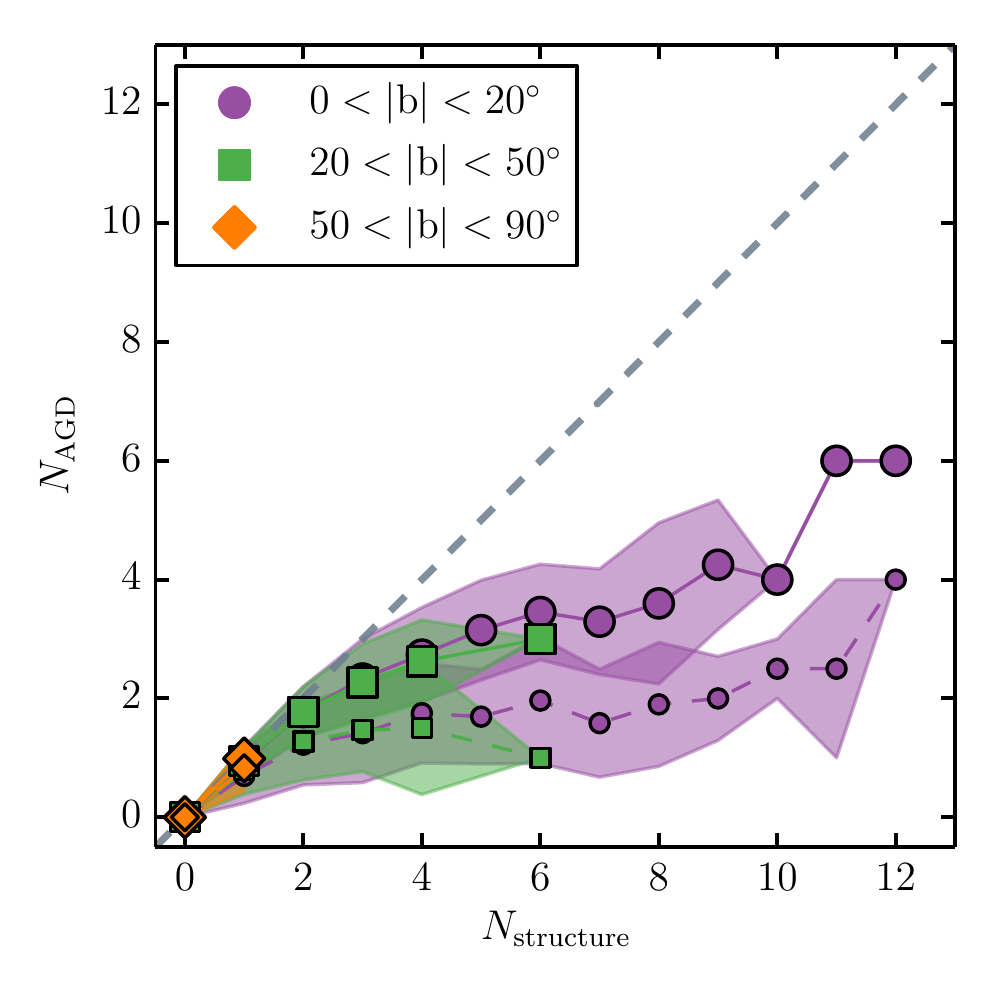}
   \caption{The number of gas structures defined by $n/T_s > 0.002\rm\,cm^{-3}/K$ along an
   LOS ($N_{\rm structure}$) compared with the number of AGD-fitted features in the synthetic line profile ($N_{\rm AGD}$). Large symbols show $N_{\rm AGD}$ matched in absorption according to Equations~\ref{e-sigs2} and~\ref{e-fwhm2}, and small symbols show $N_{\rm AGD}$ matched in absorption and emission according to Equations~\ref{e-sigs2},~\ref{e-fwhm2},~\ref{sigs_away} and~\ref{fwhm_factor}. Symbols and shading indicate the mean and standard deviations of $N_{\rm AGD}$ for each unique $N_{\rm structure}$. The samples are binned by latitude according to the inset legend.}
   \vspace{10pt}
      \label{f-cloud-matched}
\end{figure}

Given that $\tau \propto n/T_s$, we define simulated gas structures 
by selecting peaks in $n/T_s$ along each LOS. To select a threshold value, we consider the parameters of the simulated ISM
from KOK13. The gas temperature and density PDFs in their Figure 8 display strong bi-modality indicative
of multiple \hi\ phases. 
The ratio $n/T_s$ is high for the CNM and low for the WNM.  In identifying gas ``structures" along the LOS, 
we wish to mark concentrations using CNM-like peaks. We select $n\sim2\rm\, cm^{-3}$ and $T \sim 10^3\rm\, K$ as representative
values between the peaks of the published bi-modal PDFs (KOK13). These values correspond to a threshold of $(n/T_s)_{\rm thresh} = 0.002\rm\,cm^{-3}/K$. We experimented with different values of this threshold, and the subsequent results
do not change significantly. 

In Figure~\ref{f-cloud-matched-ex}, we display $(n/T_s)(s)$ (left), $\tau(v)$ (middle) and $T_B(v)$ (right) for 5
example LOS from KOK14. The positions of peaks above $(n/T_s)_{\rm thresh} = 0.002\rm\,cm^{-3}/K$ (``structures") are plotted 
as colored circles. Across all \numberkok{} synthetic LOS, there are 7582 structures with $(n/T_s)>0.002\rm\,cm^{-3}/K$.

To compare the properties of simulated gas structures with synthetic spectral lines, we first 
determine the position and velocity range (i.e. line width) of each physical gas structure.
We estimate the velocity of each gas structure, $v_{\rm sim}$, by computing the average velocity ($v(s)$)
of channels spanned by each peak in $n/T_s$ weighted by their densities ($n(s)$), specifically,

\begin{equation}
v_{\rm sim} = \frac{\int_{\rm structure} n(s)\, v(s) \, ds}{\int_{\rm structure} n(s)\,  ds},
\end{equation}  

\noindent where ``structure" refers to all pixels spanned by each peak above $(n/T_s)_{\rm thresh}=0.002\rm\,cm^{-3}/K$.  
These values are plotted in the middle and right columns of Figure~\ref{f-cloud-matched-ex}
as circles with colors corresponding to the labels in the left column ($n/T_s$).
Next, we estimate the FWHM of the structure based on its thermal and turbulent properties.
We compute the inferred thermal line width, $\Delta v_{\rm therm}$, by solving \citep[e.g., Eq. 9.31,][]{draine2011}, 

\begin{equation}
\Delta v_{\rm therm} = 2.15\,  \sqrt{\frac{T_{\rm mean} / 100\rm \, K}{M/m_{\rm H}}} = 0.190\,  \sqrt{T_{\rm mean} } ~ \rm km\,s^{-1}
\label{e-temp}
\end{equation}

\noindent where we assume $M= \mu\, m_{\rm H}$ for mean molecular weight $\mu=1.27$ (c.f., KOK13), and $T_{\rm mean}$ is the harmonic mean kinetic temperature of the gas spanned by each peak, given by,

\begin{equation}
T_{\rm mean} = \frac{\int_{\rm structure} n(s) \, ds}{\int_{\rm structure} (n/T) (s)\,  ds}. 
\end{equation} 

\noindent We then estimate the contribution from turbulent line broadening, $\Delta v_{\rm turb}$, by
computing the standard deviation of the velocities spanned by each structure, multiplied
by a factor of $2.355$ to convert to a FWHM, or,

\begin{equation}
\Delta v_{\rm turb} = 2.355 \sqrt{ {\frac{\int_{\rm structure} n(s) \, (v(s) - v_{\rm sim})^2\, ds}{\int_{\rm structure} n(s)\, ds}} } .
\label{dv_turb}
\end{equation}

\noindent The final estimate of the velocity FWHM
of each structure, $\Delta v_{\rm sim}$, is a quadratic sum of the thermal and turbulent
contributions (Equations~\ref{e-temp} and~\ref{dv_turb}), or $\Delta v_{\rm sim} = \sqrt{\Delta v_{\rm therm}^2 + \Delta v_{\rm turb}^2}$,  and has values between $\sim1-10\rm\,km\,s^{-1}$ for a range of density threshold choices. 

\subsection{Matching gas structures with \hi\ absorption lines}

To match AGD-fitted Gaussian absorption lines with gas structures along each LOS, we use two matching criteria. 
First, we define $\delta_v$ to be the difference in mean velocity of a Gaussian component fitted by AGD in absorption ($v_0$)
with the estimated velocity of the gas structure ($v_{\rm sim}$) in terms of the measured FWHM from the AGD fit ($\Delta v_0$), or,

\begin{equation}
\delta_{v} \equiv \frac{ |v_0 - v_{\rm sim}|}{\Delta v_0/2.355},
\end{equation}

\noindent For a gas structure to match a Gaussian component, we require that their positions in 
velocity be less than one standard deviation away from each other, so that,
\begin{equation}  
\delta_{v} \leq 1.
\label{e-sigs2}
\end{equation}

Second, we define $R_{\rm FWHM}$ to be the ratio of
the FWHM of a component in absorption ($\Delta v_0$) and the estimated FWHM
of the gas structure in velocity ($\Delta v_{\rm sim}$), or, 

\begin{equation}
R_{\rm FWHM} \equiv \Delta v_{\rm sim} / \Delta v_0.
\end{equation}

\noindent For a structure to match a Gaussian component, we require that the structure's 
simulated velocity FWHM ($\Delta v_{\rm sim}$), including the thermal and turbulent contributions, be similar to 
the FWHM of the Gaussian component (within a factor of 3), or,

\begin{equation}  
0.3 \leq R_{\rm FWHM} \leq 3. 
\label{e-fwhm2}
\end{equation}

\noindent We note that choices of cutoff values for $R_{\rm FWHM}$ 
does not significantly change the results, as the criterion described by Equation~\ref{e-sigs2}
dominates the matching. In addition, we emphasize that the matching criteria 
were designed to be as simple as possible to minimize imposed selection biases.

In Figure~\ref{f-cloud-matched-ex}, the circle markers for structures which match with AGD absorption components
according to Equations~\ref{e-sigs2} and~\ref{e-fwhm2}, are filled, and the matching AGD absorption component is 
plotted in the corresponding color in the middle panel. If a structure does not have an AGD match, the circle marker is unfilled.
Of the 7582 total structures, there are 6097 structures with matches to AGD absorption components.

\subsubsection{Matching \hi\ absorption lines with \hi\ emission lines}

Both \hi\ absorption ($\tau(v)$) and emission ($T_{\rm B}(v)$) information are required to constrain 
the spin temperature ($T_s$) and column density ($N(\rm H\textsc{i})$) of neutral gas in the ISM via radiative transfer.
Therefore, to compare the density and temperature of simulated gas structures with properties inferred from
observations, we need to determine the optical depth \emph{and} brightness temperature of each structure.

To match \hi\ absorption lines with \hi\ emission lines fitted by AGD, we apply a similar set of 
match criteria as described by Equations~\ref{e-sigs2} and~\ref{e-fwhm2}. Specifically, 

\begin{eqnarray}
\delta_{v, \rm AGD} & \equiv & \frac{ |v_0 - v_{0,\rm \,em}|}{\Delta v_0/2.355} \leq 1, \label{sigs_away} \\
1 \leq R_{\rm FWHM, \rm AGD} & \equiv & \Delta v_{0,\rm em} / \Delta v_0 \leq 3,
\label{fwhm_factor}
\end{eqnarray}

\noindent where ($v_{0,\rm \, em} , \, \Delta v_{0,\rm em}$) are the mean velocity and FWHM of an AGD component fitted to $T_B(v)$.
We impose the requirement that $R_{\rm FWHM,\rm AGD} \geq 1$ to 
ensure that the matched line width is larger in emission than absorption, and impose
$R_{\rm FWHM,\rm AGD}\leq3$ in order to ensure that the line widths are reasonably similar. 
We do not include a criterion for matching component amplitudes here, because
the amplitude of an absorption feature in emission is a determined by both its optical depth
and its spin temperature, which are difficult to disentangle. Furthermore, we are interested
in analyzing how well this simple approach recovers the spin temperatures of structures, 
which we analyze in Section 5.4.1, and therefore we do not impose any requirement
that the component amplitudes match at this stage. 

In Figure~\ref{f-cloud-matched-ex}, if a structure (left panel) matches with an AGD absorption line (middle panel)
and also matches with an AGD emission line (right panel) according to Equations~\ref{sigs_away} and~\ref{fwhm_factor}, the circle marker
contains a white star and the matching emission line is plotted in the corresponding color (right panel).
Of the 6097 structures with AGD absorption line matches, 4228 also
have a match to an AGD emission line.

\subsection{Quantifying Match Completeness}
\label{s:complete}

For most examples in Figure~\ref{f-cloud-matched-ex}, gas structures (left) are accounted for by the majority of the 
total optical depth along the LOS (middle). This suggests that the AGD absorption lines can be mapped to real structures. 
However, in the presence of strong line blending, as shown by the third row of Figure~\ref{f-cloud-matched-ex} (e.g., case $\rm i4952$), 
several absorption lines have nearly the same central velocity and the majority of the absorption feature cannot be matched.
Although the majority  ($90\%$) of LOS in KOK14 have $<2$ fitted components, we selected
the examples in Figure~\ref{f-cloud-matched-ex} to illustrate a range in complexity for our fitting and matching process.
In particular, the more complex LOS shown in the bottom three rows of Figure~\ref{f-cloud-matched-ex},
with $>2$ fitted components, are likely more representative of real observations. 
We discuss this issue further in Section 6.2.

In Figure~\ref{f-cloud-matched} we compare the number of structures ($N_{\rm structure}$) along
each LOS with the number of matched AGD-fitted components of the synthetic line profile ($N_{\rm AGD}$).
For each unique value of $N_{\rm structure}$, the mean (symbol) and standard deviation (shading) of $N_{\rm AGD}$
are shown. Furthermore, we break the sample into latitude bins according to the inset legend. 
Large symbols indicate the number of AGD absorption matches according Equations~\ref{e-sigs2} and~\ref{e-fwhm2}, 
and small symbols indicate the number of AGD absorption \emph{and} emission matches following 
the subsequent application of Equations~\ref{sigs_away} and~\ref{fwhm_factor}. 

Using Figure~\ref{f-cloud-matched}, we quantify the completeness ($C$) of recovery by,

\begin{equation}
C = \frac{\sum{N_{\rm AGD}}}{\sum{N_{\rm structure}}}.
\label{e:complete}
\end{equation}

\noindent For the matches between gas structures and absorption lines only (large symbols), the completeness
for the low ($0<|b|<20^{\circ}$), mid ($20<|b|<50^{\circ}$) and high latitude ($|b|>50^{\circ}$)
bins are $C_{\rm absorption}= (0.53,\, 0.67, \,0.99)$ respectively. The recovery is best at the highest latitudes
where the number of gas structures and AGD components are smallest and the 
LOS complexity is minimized, allowing for simple and robust AGD fits.
At low latitudes, the blending of gas velocities and AGD components 
makes it difficult to associate unambiguous spectral components with 
$(n/T_s)$ peaks.

When matching gas structures to \hi\ emission instead of absorption, the recovery completeness 
is $C_{\rm emission} = (0.46, \,0.53,\, 0.93)$ for low, mid and high latitudes
respectively. The completeness is worse in emission than absorption. As observed
in Figure~\ref{f-cloud-matched}, broad components associated with high-temperature \hi\ 
are prominent, thereby making the match to corresponding gas structures
more difficult. Although Gaussian analysis has been used extensively in the past to identify gas 
populations in local and external galaxies, 
this is among the first statistical quantification of the correspondence between Gaussian \hi\ emission
components and individual gas structures. 

When matching structures to both absorption and emission (small symbols in Figure~\ref{f-cloud-matched}), the completeness
for the low, mid and high latitude bins are: $C_{\rm both}= (0.29,\, 0.38, \,0.83)$ respectively. 
At all latitudes, the structure recovery completeness declines when the
match between absorption and emission is performed. 
The bottom row of Figure~\ref{f-cloud-matched-ex} displays an example
of this decline. Whereas 4/5 structures along the LOS are recovered by AGD absorption components,
only 2/4 of those absorption components have matches in emission according to 
Equations~\ref{sigs_away} and~\ref{fwhm_factor}. 
The structures selected by the matching process with both absorption and emission
are biased towards unambiguous features in all three spaces. 
Nevertheless, the completeness of Gaussian decomposition for the multi-phase structures
seen in both emission and absorption is good at $|b|>50^{\circ}$.
This is certainly promising for future large data sets -- even with $\sim50\%$ 
attrition a large number of structures can be retrieved over a wide range
of interstellar conditions.

As discussed in the context of Figure~\ref{f-cloud-matched-ex}, in some cases the 
fraction of total absorption or total emission that can be accounted for by gas structures along
the LOS is low. This is especially true when $N_{\rm AGD}\geq2$, although those comprise the minority
of cases in KOK14. In the future, improving our selection method for gas 
structures along the LOS beyond a single cutoff value in ($n/T_s$) will improve this completeness of structure recovery. 
In addition, developing additional criteria for structure-component matching based on their amplitudes
and/or total column densities will enable us to better quantify the range of gas structures that
can be recovered reliably by fitted spectral lines. The analysis presented here represents a first step
in our ongoing investigation. 

 \begin{figure}
 \vspace{-0.2in}
   \includegraphics[width=0.45\textwidth]{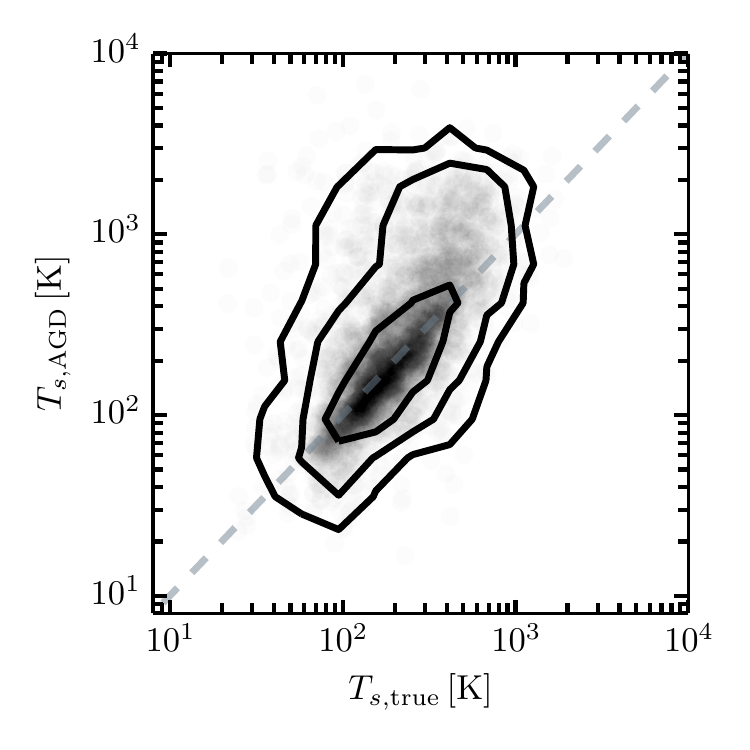}
   \vspace{-0.1in}
   \caption{``True" simulated spin temperature ($T_{s,\rm true}$, Equation~\ref{tstrue}) versus inferred spin temperature ($T_{s, \rm AGD}$, Equation~\ref{ts-agd}) for all structures which match fitted absorption and emission lines according to Equations~\ref{e-sigs2},~\ref{e-fwhm2},~\ref{sigs_away} and~\ref{fwhm_factor}. Contours indicate the $1,\,2\, \rm and\, 3\sigma$ limits.}
      \label{f-temp-compare}
\end{figure}

\subsection{Observed vs. ``True" Gas Properties}
\label{s:clouds}

Given a sample of gas structures with matches to AGD components in 
absorption and emission, we compare the true temperatures in the simulation
with the values inferred from AGD-fitting of the spectra.

\begin{figure*}
   \centering
  \includegraphics[width=\textwidth]{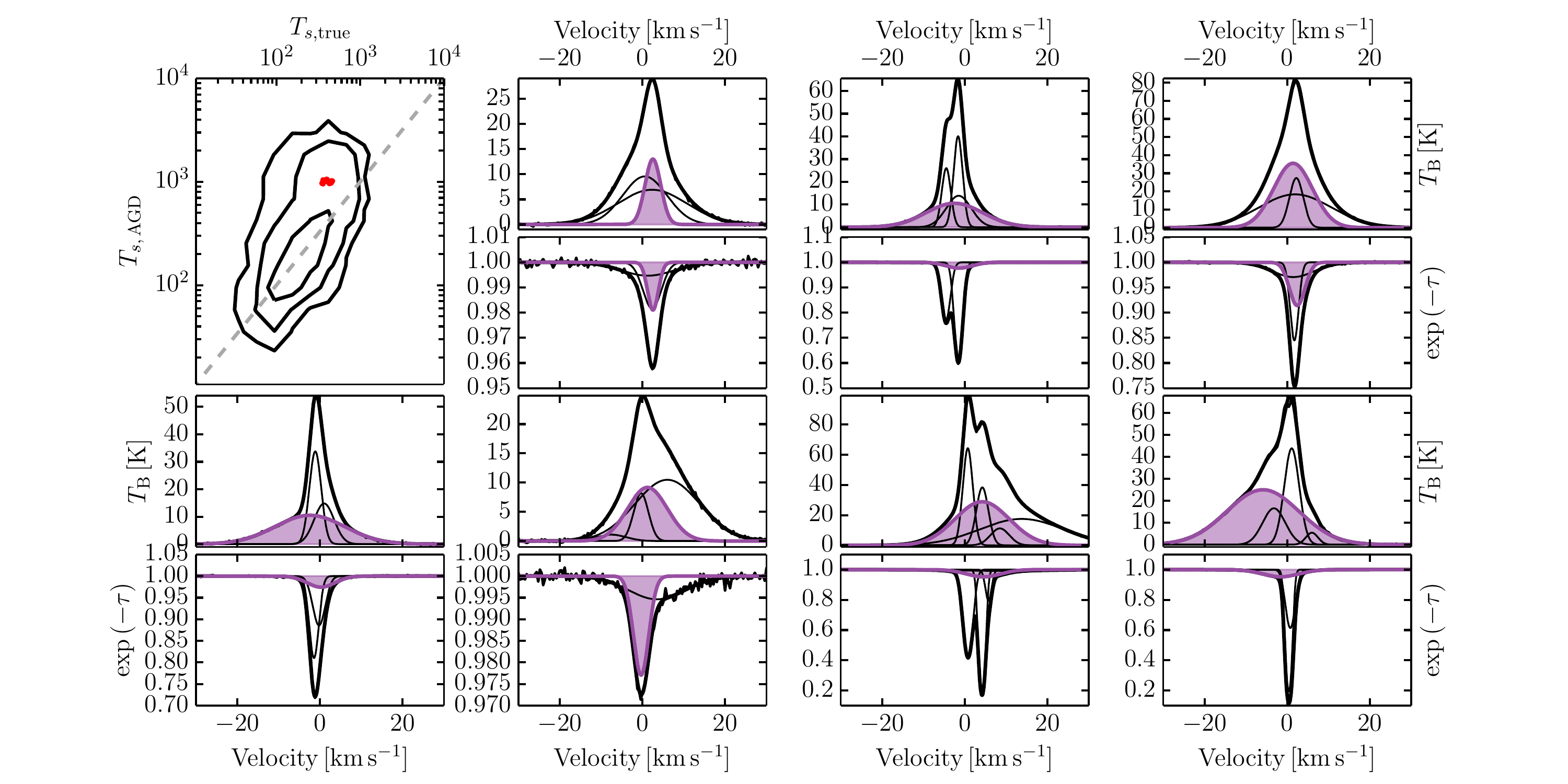}
   \caption{Example synthetic \hi\ emission and absorption spectral pairs in which the AGD-derived spin temperature ($T_{s,\rm AGD}$) overestimates the simulated spin temperature ($ T_{s,\rm true}$). The top-left panel is reproduces Figure~\ref{f-temp-compare}, with 7 examples around $T_{s,\rm true}=400\rm\,K$ and $T_{s,\rm AGD}=1000\rm\,K$ highlighted in red. The matched \hi\ absorption and emission pairs corresponding to these highlighted points to are plotted and shaded in purple in the accompanying panels, along with the full AGD decomposition for each spectrum (black). }
               \label{f-inspect}
\end{figure*}

\subsubsection{Spin Temperature}
\label{s-spintemp}

With the goal of estimating spin temperature automatically for
a large number of spectra, we take a more simplified approach than what has
been done in HT03 or \citet{murray2015}.
For each AGD match between absorption and emission,
we start with the isothermal spin temperature as a function of velocity, $T_{s, \rm AGD}(v)$, 
given by,

\begin{equation}
T_{s,\rm AGD}(v) = \frac{T_{B,\rm AGD}(v) }{1- e^{- \tau_{\rm AGD}(v)}},
\label{ts-perch}
\end{equation}

\noindent where $T_{B,\rm AGD}(v)$ and $\tau_{\rm AGD} (v)$ are the matched set of Gaussian functions
fitted by AGD to $T_B(v)$ and $\tau(v)$ respectively. 
This method assumes a single temperature gas within each velocity channel. 
To estimate average spin temperature per AGD component,
we compute the optical depth-weighted spin temperature per component,

\begin{equation}
T_{s, \rm AGD} \equiv \frac{\int \tau_{\rm AGD}(v)\, T_{s,\rm AGD}(v) dv}{\int \tau_{\rm AGD}(v) dv}.
\label{ts-agd}
\end{equation}

This approach produces a weighted mean temperature for each component, given 
that $T_{s,\rm AGD}(v)$ is smaller near the peak of $\tau_{\rm AGD}(v)$ and larger 
away from it. We note that there are several possible ways to estimate mean temperature from $\tau(v)$ and $T_B (v)$ observations,
and some discussion of the pros and cons of each method are given in HT03 and \citet{dickey2003}. 
It is important to note that Equation~\ref{ts-agd} works well if, within a multi-phase structure, the CNM and WNM
are centered at a similar radial velocity. However, if the CNM
is shifted in velocity relative to the WNM due to turbulent motions, so that the peaks of $T_{B,\rm AGD}(v)$
and $\tau_{\rm AGD}(v)$ are slightly offset, Equation~\ref{ts-agd} will overestimate $T_{s, \rm AGD}$.
HT03 and \citet{murray2015} have allowed for the CNM motion relative to 
the WNM in their temperature estimates by using a more complex fitting approach 
where spin temperature is fitted simultaneously with all WNM components.
This is, however, computationally expensive for us to implement at this stage.

To estimate the temperature of a simulated gas structure, we compute the harmonic mean 
temperature, $T_{s, \rm true}$, within the pixels spanned by each peak in $n/T_s$-space. 
Specifically, 

\begin{equation}
T_{s,\rm true} = \frac{\int_{\rm structure} n (s) ds}{\int_{\rm structure} (n/T_s) (s) ds}.
\label{tstrue}
\end{equation}  

In Figure~\ref{f-temp-compare}, we compare the simulated spin temperature with the 
inferred spin temperature derived using our radiative transfer approach
for all structures with AGD matches in absorption and emission according 
to Equations~\ref{e-sigs2},~\ref{e-fwhm2},~\ref{sigs_away} and~\ref{fwhm_factor}. 
We recover nearly the full range of spin temperatures found in KOK14,
which indicates that our structure selection method is likely not missing 
a significant gas population.
Furthermore, the AGD and true estimates agree
within the $1\sigma$ contours of Figure~\ref{f-temp-compare}.
However, at high temperatures, where $T_{s, \rm true}> 400\rm\,K$, the 
AGD temperature overestimates the true spin temperature.

 \begin{figure}[b!]
     \includegraphics[width=0.5\textwidth]{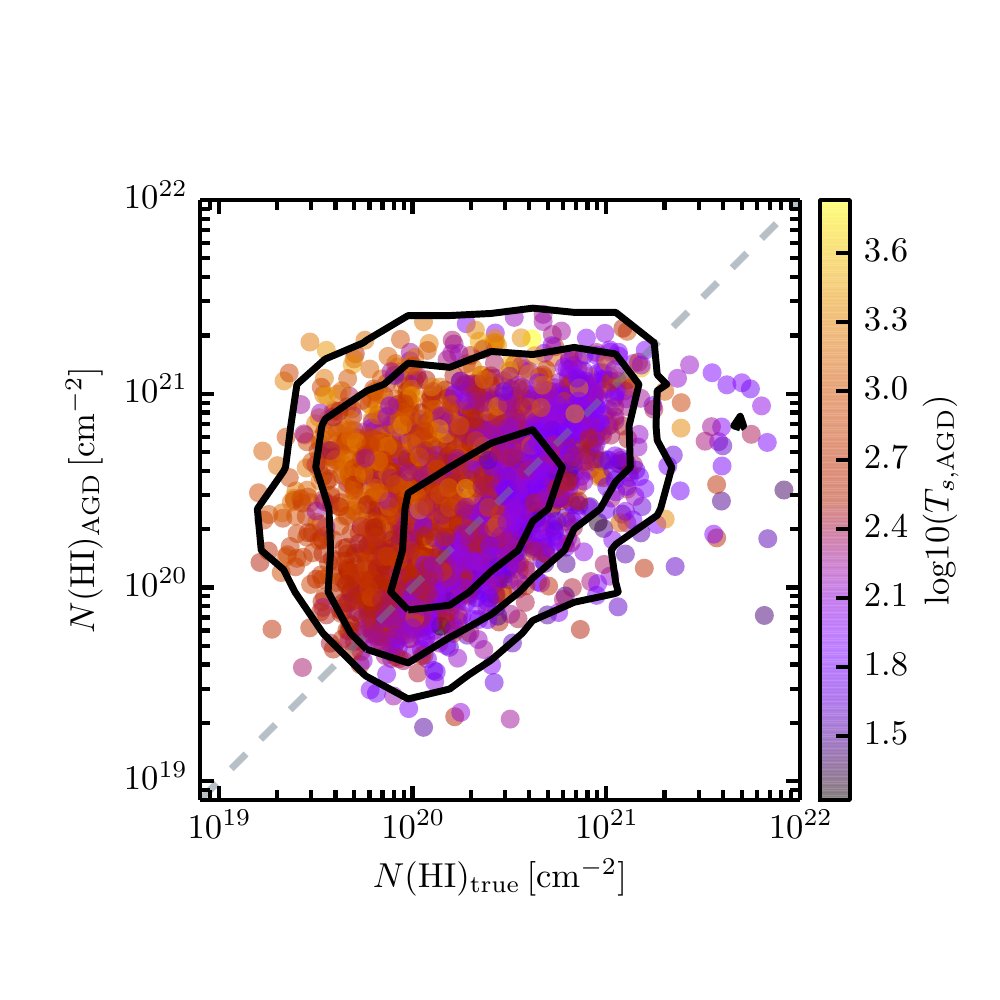}
     \vspace{-0.4in}
   \caption{``True", simulated column density ($N({\rm H\textsc{i}})_{\rm true}$, Equation~\ref{nhtrue}), versus ``observed" column density ($N({\rm H\textsc{i}})_{\rm AGD}$, Equation~\ref{nhcomp}) for all peaks which match fitted absorption and emission components according to Equations~\ref{e-sigs2},~\ref{e-fwhm2},~\ref{sigs_away} and~\ref{fwhm_factor}. Contours indicate the $1,\,2\, \rm and\, 3\sigma$ limits.}
   \vspace{10pt}
      \label{f-nh-compare}
\end{figure}

To understand why AGD overestimates spin temperature at high temperatures,
Figure~\ref{f-inspect} displays a set of example matched spectral lines with $T_{s,\rm true}\sim400\rm\,K$ 
and $T_{s,\rm AGD}\sim1000\rm\,K$. We reproduce the contours from 
Figure~\ref{f-temp-compare} in the top left panel of Figure~\ref{f-inspect}, and include the corresponding matched
\hi\ emission and absorption components in the accompanying panels, highlighted with purple shading. 
In all highlighted cases, the emission component is slightly offset in velocity from the 
corresponding absorption component. 

\begin{figure*}
   \includegraphics[width=\textwidth]{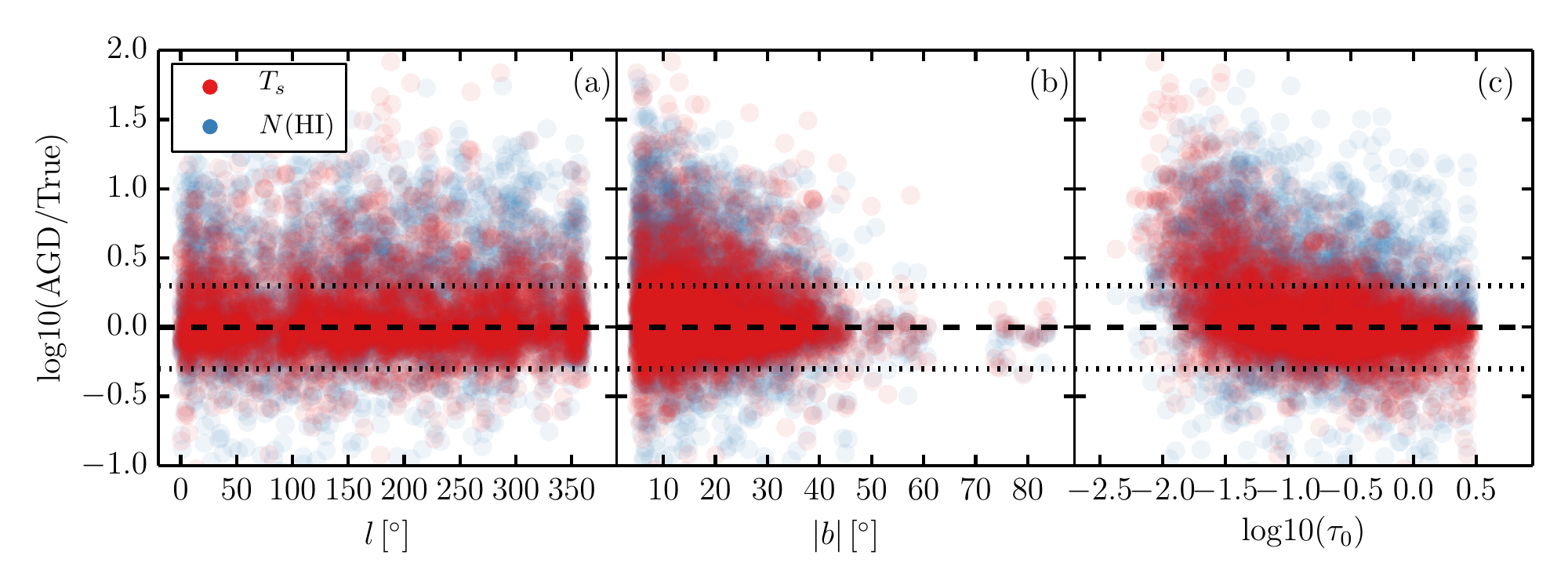}
   \caption{The ratio of inferred (``AGD") to simulated (``True") properties of gas structures matched with Gaussian spectral lines fitted to \hi\ emission and absorption lines as a function of various LOS parameters. (a): Galactic longitude, $l\,[^{\circ}]$; (b): absolute Galactic latitude, $|b|\,[^{\circ}]$; (c): peak optical depth of the matched absorption line, $\tau_0$; Spin temperature, $T_{s,\rm AGD}/T_{s,\rm true}$ (red) and column density, $N({\rm H\textsc{i}})_{\rm AGD}/N({\rm H\textsc{i}})_{\rm true}$ (blue). Dotted lines indicate factors of 2. }
   \vspace{10pt}
      \label{f-analyze}
\end{figure*}

As discussed above, this offset is caused by interstellar turbulence.
When Equation~\ref{ts-agd} is applied, the resulting
spin temperature will be overestimated if we do not account for this velocity offset. 
For example, if we estimate 
the spin temperature using the peak brightness temperature and peak optical depth of 
highlighted components in Figure~\ref{f-inspect}, 
we get a value that agrees much more closely with $T_{s,\rm true}$, since  $T_{B,\rm peak}/(1-e^{-\tau_{\rm peak}}) \sim 400\rm\,K$.
A more complex radiative transfer treatment such as the method of HT03
and \citet{murray2015}, accounts for this effect, which is strongest for those components with the
highest turbulent velocity offset between the CNM and WNM. 
We find that $T_{s,\rm AGD}$ most strongly over-estimates $T_{s,\rm true}$ when 
the velocity offset between absorption and emission is highest. However, most components are not affected. 
We will fine-tune our radiative transfer treatment in future work. 

\subsubsection{Column Density}

The \hi\ column density ($N(\rm H\textsc{i}$)) is given by,

\begin{equation}
N({\rm H\textsc{i}}) = C_0\, \int T_s(v)\, \tau(v)\, dv,
\end{equation}

\noindent where $C_0=1.813\times10^{18}\rm\,cm^{-2}\,K^{-1}\,(km\,s^{-1})$. 
For a pair of matched AGD absorption and emission lines, we  
compute the column density per component ($N(\rm H\textsc{i})_{\rm AGD}$) as,

\begin{equation}
N({\rm H\textsc{i}})_{\rm AGD} = C_0  \, T_{s,\rm \, AGD} \int \tau_{\rm AGD}(v) \,dv,
\label{nhcomp}
\end{equation}

\noindent where $T_{s,\rm \, AGD}$ is computed using Equation~\ref{ts-agd}
and $\tau_{\rm AGD}(v)$ is the Gaussian function fitted by AGD to $\tau(v)$. 

The simulated column density of each gas structure is given by,

\begin{equation}
N({\rm H\textsc{i}})_{\rm true} = \int_{\rm structure} n(s) \, ds \, .
\label{nhtrue}
\end{equation}

In Figure~\ref{f-nh-compare}, we compare the ``true" and inferred column density for all structures with AGD matches in absorption and emission according to Equations~\ref{e-sigs2},~\ref{e-fwhm2},~\ref{sigs_away} and~\ref{fwhm_factor}. 
As in Figure~\ref{f-temp-compare}, the AGD and true estimates agree 
within the $1\sigma$ contours. However, outside the $1\sigma$ contours, $N({\rm H\textsc{i}})_{\rm AGD}$ 
overestimates $N({\rm H\textsc{i}})_{\rm true}$, in part because the uncertainty in $T_{s,\rm AGD}$ is propagated to $N({\rm H\textsc{i}})$ via Equation~\ref{nhcomp}. We color the points in Figure~\ref{f-nh-compare} by $T_{s,\rm AGD}$ to illustrate this.  The most discrepant points correspond to the highest values of $T_{s,\rm AGD}$, where the uncertainty in the matching process is highest (c.f., Figure~\ref{f-temp-compare}).

In Figure~\ref{f-analyze}, we investigate the scatter present in 
Figures~\ref{f-temp-compare} and~\ref{f-nh-compare} by plotting the ratios of the inferred (``AGD") to direct (``true")
estimates of spin temperature and column density as a function of various LOS parameters.
These include Galactic longitude ($l$; a),
absolute Galactic latitude ($|b|$; b), and the peak optical depth of the matched AGD line ($\tau_0$; c).
In each panel, the data points are colored according to $T_{s,\rm AGD}/T_{s, \rm true}$ (red) 
and $N({\rm H\textsc{i}})_{\rm AGD}/N({\rm H\textsc{i}})_{\rm true}$ (blue). 

In Figure~\ref{f-analyze}, the ratio of AGD to true spin temperature and column density falls within a factor of 2 for the majority ($68\% , \,1\sigma$ contours) of structures at all longitudes (a), latitudes (b) and peak optical depths (c) probed. This indicates that the AGD method is able to recover these properties reasonably well. The scatter in the ratio of AGD to true column density is also larger than for spin temperature, due to the fact that any uncertainty in $T_{s,\rm AGD}$ is propagated to $N({\rm H\textsc{i}})_{\rm AGD}$ via Equation~\ref{nhcomp}. 

The scatter in AGD/True appears to be constant with longitude ($l$), yet increases at low latitudes ($|b|$) and low peak optical depths ($\tau_0<0.1$).
This is caused by the increase in LOS complexity at low latitudes, whose effect was noted in Figure~\ref{f-cloud-matched}, and the increased likelihood for a low-$\tau_0$ component to be affected by velocity blending. Examples of strong line blending are shown in Figure~\ref{f-temp-compare}.

However, the agreement between AGD and true spin temperature
and column density for the majority of simulated gas structures given our simple 
structure selection and matching prescription is encouraging,
and indicates that automatic routines for identifying and analyzing spectral components from \hi\ observations -- essential for future large 
observed and simulated datasets -- can be successful
in recovering true properties for a large fraction of interstellar gas structures. 

\section{Comparing Real and Synthetic \hi\ Spectra }
\label{s:agd_compare}

After analyzing the biases of Gaussian analysis in recovering gas structures
and their properties using synthetically spectra from simulations, 
we proceed to compare observed (21-SPONGE) 
and simulated (KOK14) \hi\ spectra via AGD-fitted Gaussian parameters.

\begin{figure}
   \centering
  \includegraphics[width=0.5\textwidth]{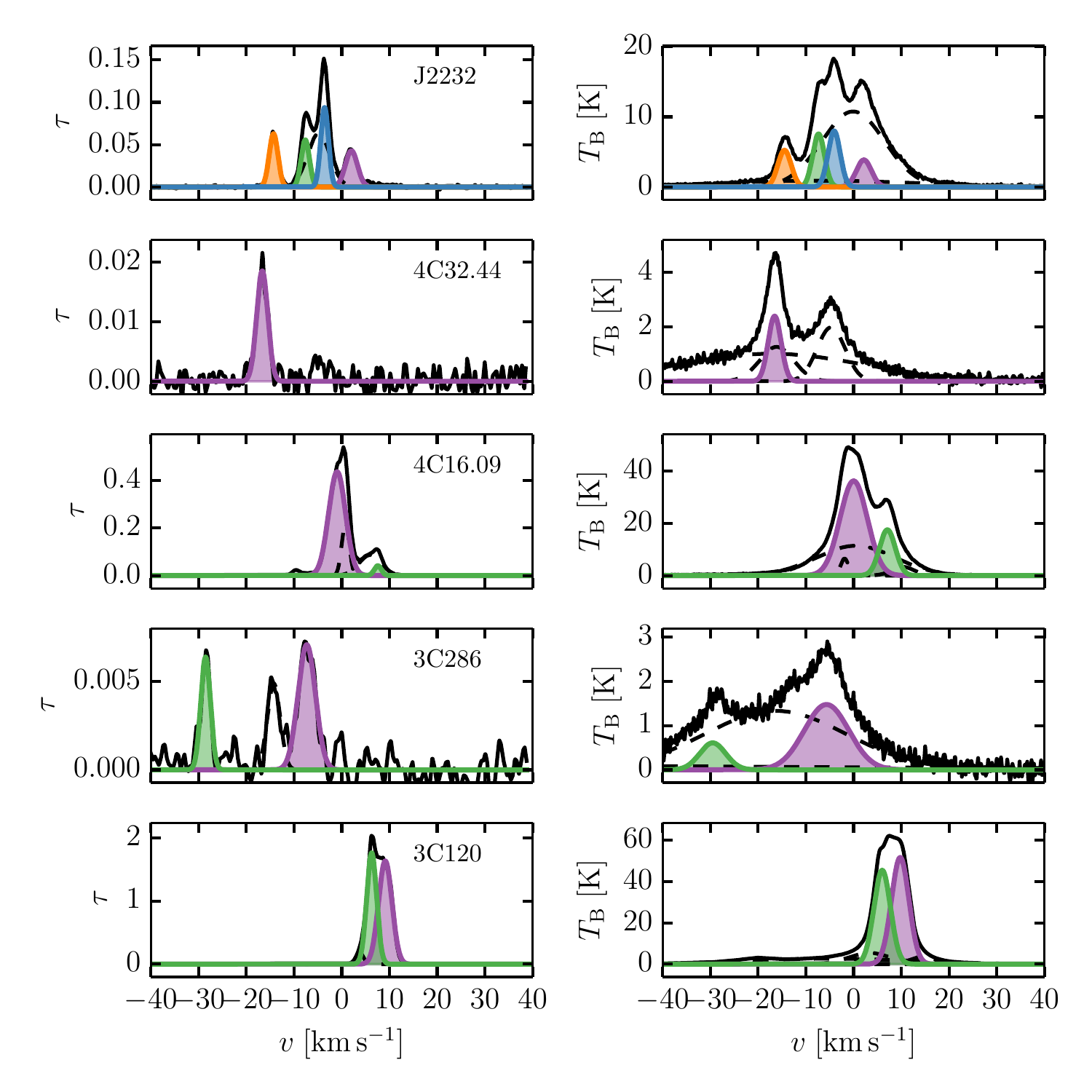}
   \caption{Example LOS \hi\ absorption spectra (left) and corresponding
   \hi\ emission spectra (right) from 21-SPONGE, with AGD-fitted Gaussian decompositions overlaid (dashed black). 
   Components which match between absorption (left) and emission (right) according to Equations~\ref{sigs_away} and~\ref{fwhm_factor} are 
   indicated by matching colors between the left and right columns.}
               \label{f-los-sp}
\end{figure}

With the AGD decompositions of \hi\ emission and absorption for 
all 21-SPONGE and KOK14 LOS, we apply the criteria 
described by Equations~\ref{sigs_away} and~\ref{fwhm_factor} to 
match as many AGD lines between absorption and emission as possible.
For KOK14, this does not take into account matching with gas structures 
along the LOS as described in Section~\ref{s:sim_an}, in the interest
of eliminating as many biases as possible in our comparison between 
the matching statistics of KOK14 and 21-SPONGE (which does not have 
LOS density and temperature information for defining structures). 

Figure~\ref{f-los-sp} displays the matches between \hi\ absorption 
and \hi\ emission for a set of 5 example observed LOS from 
21-SPONGE. All components fitted by AGD to each LOS are shown in 
dashed black, and the components which satisfy the matching criteria 
described by Equations~\ref{sigs_away} and~\ref{fwhm_factor} are shown in colors. 

\subsection{Number of components along LOS}

In Table~\ref{t:AGD} we list the total number 
of AGD components in each \hi\ absorption and emission dataset, as well as the mean and standard deviation
of the number of components ($N_{\rm AGD}$) per LOS.
In addition, we list the total number of matched components between absorption and emission, and the mean and standard deviation number of matches per LOS. 

\begin{figure*}
   \centering
  \includegraphics[width=0.45\textwidth]{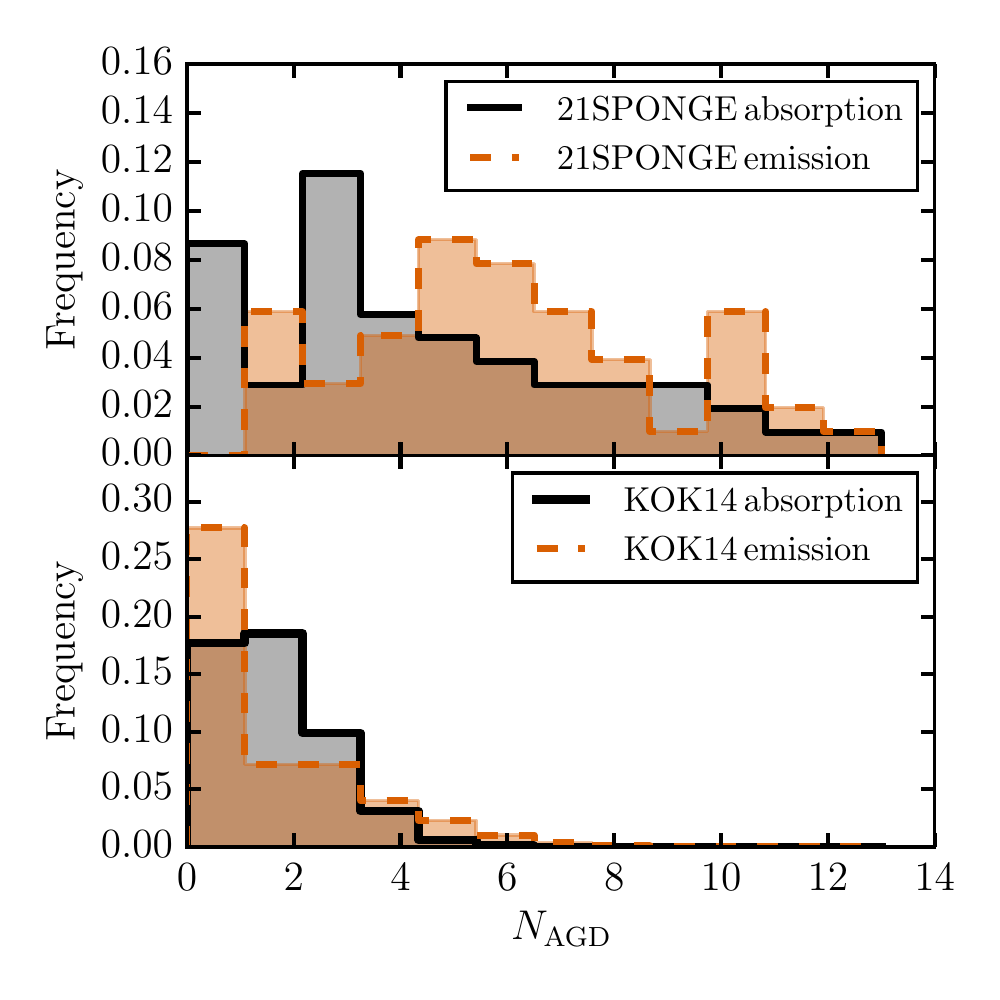}
    \includegraphics[width=0.45\textwidth]{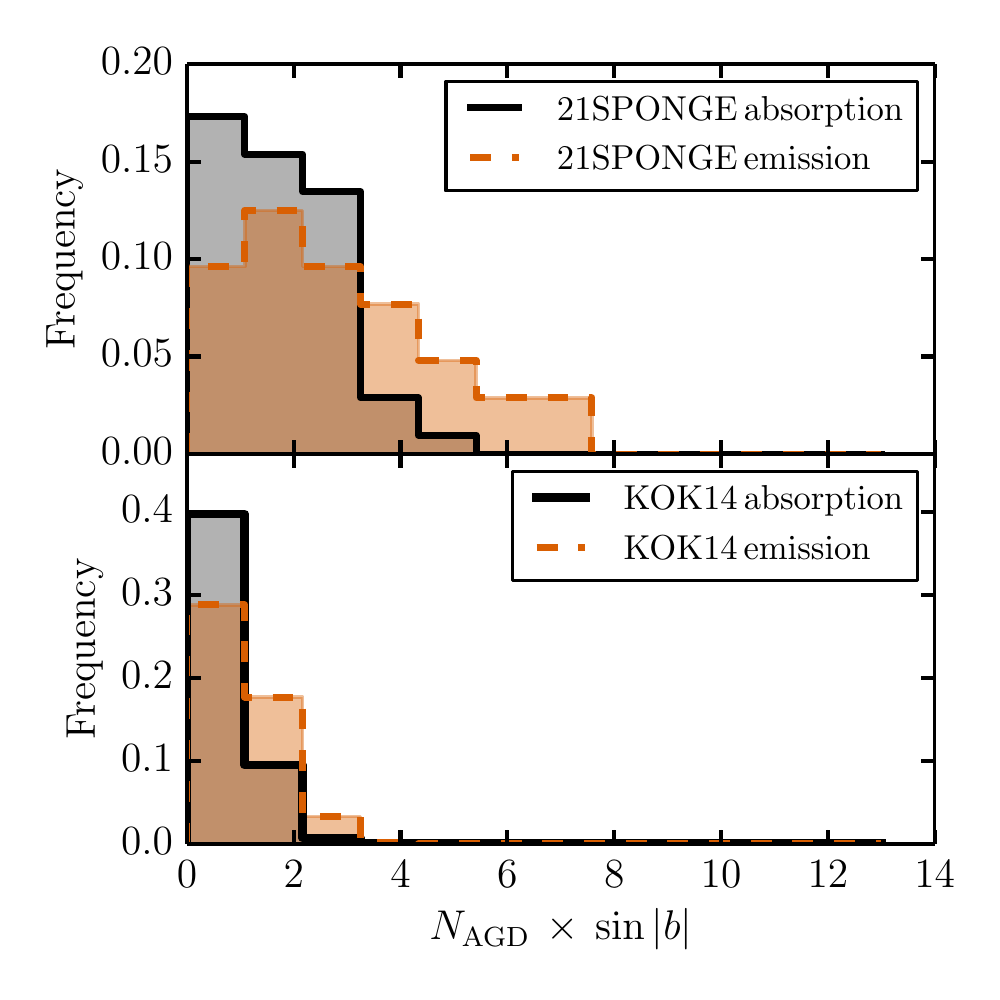}
     \vspace{-0.1in}
   \caption{Left: Histograms displaying number of AGD lines ($N_{\rm AGD}$) fit to \hi\ absorption (black) and emission (orange) observations from 21-SPONGE (top) and synthetic observations by KOK14 (bottom). Right: Histograms displaying number of AGD lines per unit path length in the ``vertical" direction ($N_{\rm AGD}\times \sin{|b|}$) for \hi\ absorption (black) and emission (orange) observations from 21-SPONGE (top) and synthetic observations by KOK14 (bottom). }
               \label{f-count}
\end{figure*}

From Table~\ref{t:AGD}, the mean value of $N_{\rm AGD}$ is 
more than a factor of two greater for the observed 21-SPONGE \hi\ absorption and emission spectra
than KOK14, despite large scatter. 
Figure~\ref{f-count} displays
histograms of $N_{\rm AGD}$ for 21-SPONGE (top left panel) and KOK14 (bottom left panel)  
absorption (black solid) and emission (orange dashed) components. In agreement with the statistics
shown in Table~\ref{t:AGD}, the maximum number of components fitted to 21-SPONGE absorption (12)
is a factor of two higher than KOK14 (6).

In the right panels of Figure~\ref{f-count}, we account for the effect or different viewing 
angles by multiplying $N_{\rm AGD}$ by $\sin{|b|}$ 
for all 21-SPONGE and KOK14 components. 
This quantity is the effective number of components in the vertical direction of the 
simulated or observed volume. Since $N_{\rm AGD} \times \sin{|b|}$ is still larger 
for the 21-SPONGE than KOK14, it suggests that the discrepancy between is not a $\sin{|b|}$ effect.

The larger number of components found in the 21-SPONGE spectra, after correcting for observing angle,
indicates that the velocity range used to produce the KOK14 synthetic spectra is smaller than what is sampled by observations.
The KOK13 simulations are known to have a relatively low vertical velocity dispersion ($\sim 5-7\rm\,km\,s^{-1}$),
somewhat smaller than observed values. The lower velocity dispersion also yields a scale
height somewhat smaller than observations. We will return to this effect in Section 6.2.2.
In more recent simulations \citep{kim2016sub} with a better
treatment for supernovae, velocity dispersions are in fact larger, and it
will be interesting to test whether this will lead to an increase in the number of 
AGD features per LOS.

We emphasize that the results shown in Figure~\ref{f-count} are derived from identical implementation of AGD to 
real and synthetic \hi\ spectra, and thus the comparison is unaffected by biases introduced in spectral line analysis. 
Therefore, although the caveats described above are known from external analysis of the KOK13 simulations,
Figure~\ref{f-count} suggests that $N_{\rm AGD}$ reflects the total velocity range and path length.

Furthermore, from Table~\ref{t:AGD}, although the number of matches per LOS is consistent with the number 
of fitted lines per LOS ($N_{\rm AGD}$) in KOK14,
there are comparatively fewer matches per LOS than $N_{\rm AGD}$ in 21-SPONGE. 
This difference likely comes from the so-called mismatch of angular resolution.
In 21-SPONGE, the angular resolution of the \hi\ absorption measurements is
determined by the size of the background source, not the telescope beam
(and is therefore $<1-40''$). However, the \hi\ emission spectrum has an angular resolution of $3.5'$.
Therefore, the \hi\ emission spectrum may not sample the same structures seen in absorption, especially if there
is significant emission structure on angular scales below the resolution limit. This mismatch complicates the matching process and causes a larger attrition rate for observations. Simulations, on the other hand, 
do not suffer from this problem, as emission and absorption are derived using the
same angular resolution. In the future, we plan to quantify this effect by smoothing simulated spectra, and we further 
compare the 21-SPONGE and KOK14 emission properties in detail in Section 6.3.

\subsection{Properties of \hi\ absorption components}
\label{abs-compare}

To compare 21-SPONGE and KOK14, 
Figure~\ref{f:gauss_abs} displays cumulative distribution functions (CDFs) of Gaussian
parameters fitted by AGD to \hi\ absorption spectra observed by 21-SPONGE (black) 
and simulated by KOK14 with (blue) and without (orange) the WF effect,
normalized by the total number of components (see Table 1).
These parameters include amplitude ($\tau_0$; left panel), FWHM ($\Delta v_0$ in $\rm km\,s^{-1}$; center panel) and 
mean velocity ($v_0$ in $\rm km\,s^{-1}$; right panel). For each dataset, we also plot the CDFs of 1000 
bootstrapped samples (shown in lighter-shaded colors according to the legend) to illustrate 
the effect of sample size and outliers on the shape of the CDF. 

\subsubsection{Comparison with previous studies}

We include the results of the by-hand Gaussian decomposition 
of the first \numberspongedr{}/\numbersponge{} 21-SPONGE sources 
(dashed purple; Murray et al.\,2015, ``DR1") and the Millennium Arecibo 21cm 
Absorption Line Survey (dashed green; HT03) in Figure~\ref{f:gauss_abs}. 
With lower sensitivity in optical depth, the HT03 distribution contains fewer $\tau_0<10^{-2}$ 
components than are found in the 21-SPONGE or KOK14 AGD decompositions. 
However, the 21-SPONGE DR1 $\tau_0$ distribution agrees very well with the 21-SPONGE 
AGD distribution, which indicates that although the AGD algorithm was trained using component 
parameters from HT03, it is successfully able to recover lower-$\tau$ amplitudes found in the 
higher-sensitivity 21-SPONGE and KOK14 spectra.  This agreement was also noted in the comparison 
between by-hand and AGD analysis of a subset of the 21-SPONGE sample (Lindner et al.\,2015). 
In addition, the 21-SPONGE AGD $\Delta v_0$ distribution agrees very well with DR1 and HT03,
indicating that for a wide range in optical depth sensitivity, a similar range in Gaussian spectral line widths can be recovered.

\subsubsection{Influence of local box simulations}

\begin{figure*}
	\vspace{-10pt}
   \includegraphics[width=\textwidth]{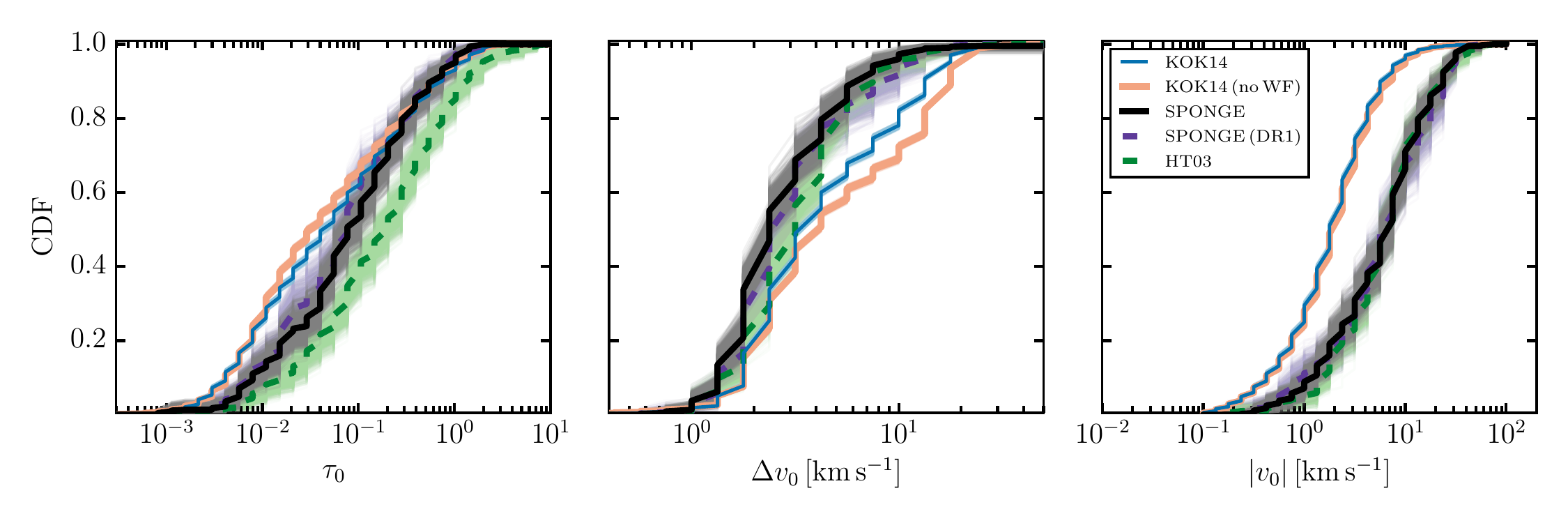}
   \caption{Cumulative distribution functions (CDFs) of Gaussian parameters, including optical depth amplitude ($\tau_0$), FWHM ($\Delta v_0$) and absolute mean velocity ($|v_0|$), of the components fitted by AGD to observed (21-SPONGE), synthetic (KOK14) \hi\ absorption spectra, including previous by-hand results from 21-SPONGE (purple, DR1; Murray et al. 2015) and HT03 (green) for comparison. For each dataset, we plot the CDFs of 1000 bootstrapped samples (shown in lighter-shaded colors according to the legend) to illustrate the effect of sample size and outliers on the CDF. }
      \label{f:gauss_abs}
\end{figure*}

From the righthand panel of Figure~\ref{f:gauss_abs}, 
the observed 21-SPONGE (AGD and DR1) and HT03 absolute mean velocities  ($|v_0|$) 
agree very well. However, the KOK14 spectra are dominated by components 
with $v_0 < 10\rm\,km\,s^{-1}$. 
This difference may be caused by
the fact that the KOK14 spectra are constructed 
with a limited path length ($s<3\rm\,kpc$) based on their local box simulations, 
in which the sources of \hi\ lines are limited by nearby gas with small variations of galactic rotation velocity.
The limited range of $v_0$ in KOK14 spectra causes more components to have similar central velocity.

\begin{figure}
             \includegraphics[width=0.5\textwidth]{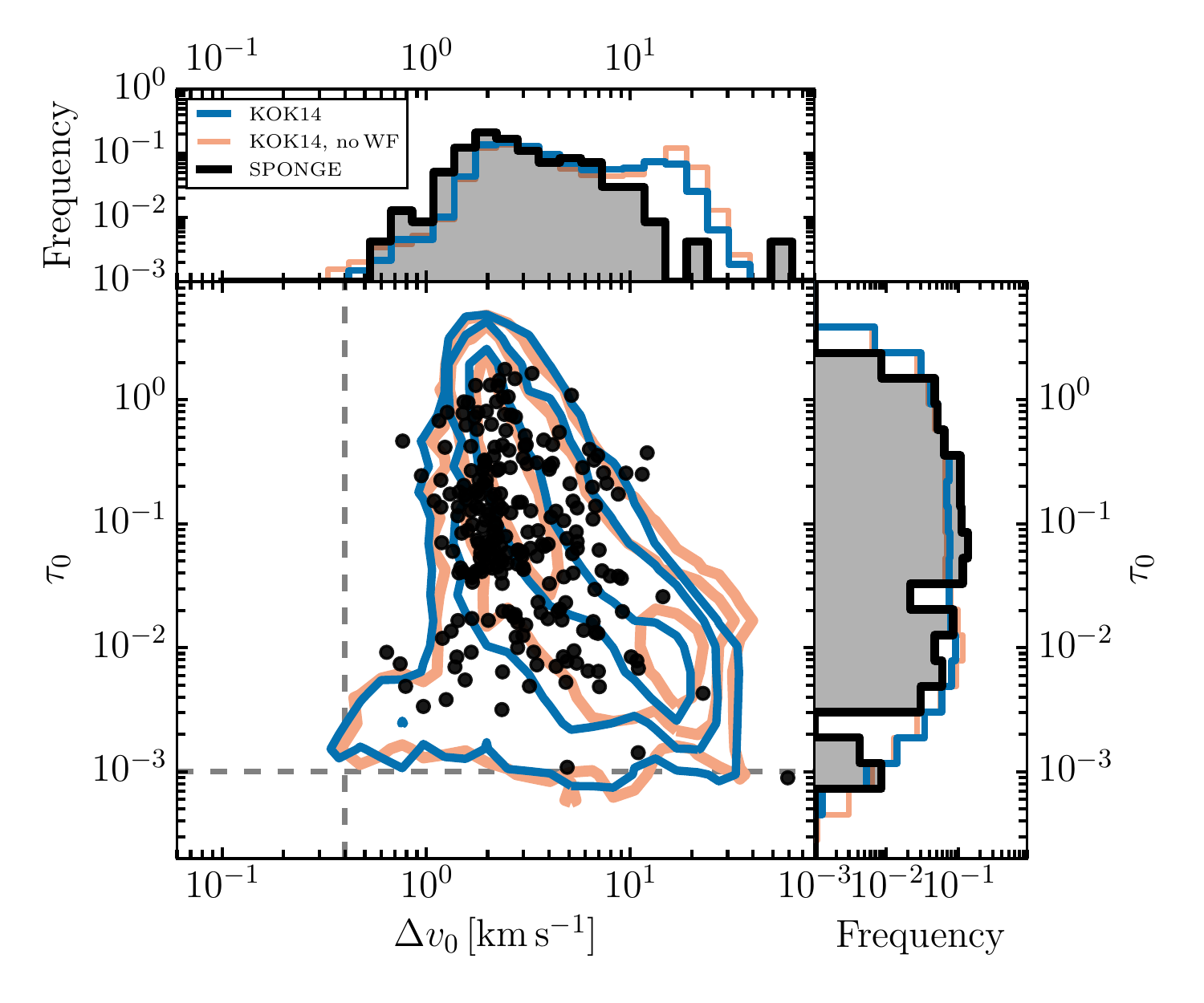}
   \caption{Parameters of Gaussian components ($\Delta v_0$, $\tau_0$) fitted by AGD to \hi\ absorption spectra from 21-SPONGE (black) and KOK14 with the WF effect (blue) and without the WF effect (orange). Contours indicate the 1, 2, and 3$\sigma$ limits for the KOK14 distributions. Marginal histograms display the same results according to the legend. }
      \label{f-cloud-hist}     
      \vspace{10pt}
\end{figure}

To test the influence of the local box and lack of global rotation effects in the simulation, 
we consider the effect of latitude on the matching statistics.
For high latitude LOS ($|b|>50^{\circ}$) in KOK14, the
number of absorption fits, emission fits and matches per LOS are consistent with the full KOK14 sample (i.e., all latitudes).
However, for 21-SPONGE, at high latitudes ($|b|>50^{\circ}$) there are $1.9\pm 1.1$ absorption fits per LOS, 
$4.4\pm 1.7$ emission fits per LOS, and $0.9 \pm 0.9$ matches per LOS, which are more consistent 
with the full KOK14 sample than with the full 21-SPONGE sample (c.f., Table~\ref{t:AGD}).
With increasing latitude, the effect of galactic rotation on spectral line properties
declines. The consistency in matching statistics between 21-SPONGE and KOK14 
at high latitudes suggests that simulated Galactic rotation dynamics play an important 
role in the differences between observed and simulated spectral line 
properties at low latitudes. Therefore, the Gaussian fitting process and match between emission and absorption 
(Equations~\ref{sigs_away} and~\ref{fwhm_factor}) in KOK14 likely suffer from more completeness issues
than in observed 21-SPONGE spectra at low latitudes. 
This also suggests that our completeness statistics in Section~\ref{s:complete} will be improved
in future simulations, in which a larger box, and more realistic supernova feedback are implemented
to push gas to higher scale heights and a wider range in velocity. 

\subsubsection{Minimum CNM temperature?}

To highlight the comparison between 21-SPONGE and KOK14 we plot $\tau_0$ vs. $\Delta v_0$  
in Figure~\ref{f-cloud-hist}, including marginal histograms of both parameters.
The median $1\sigma$ sensitivity limit in $\tau$ is indicated by the dashed horizontal line ($\sigma_{\tau}=\rm 10^{-3}$),
and the 21-SPONGE velocity resolution of $0.4\rm\,km\,s^{-1}$ is indicated by the dashed vertical line.

From the top panel of Figure~\ref{f-cloud-hist}, 
we observe a sharp cutoff in $\Delta v_0\sim1-2\rm\,km\,s^{-1}$ in 21-SPONGE and KOK14.\footnote{We note that the components with $\Delta v_0<1\rm\,km\,s^{-1}$ in 21-SPONGE and KOK14
are likely spurious fits, given that the accuracy of the AGD decomposition is known to be $80\%$ in absorption (c.f., Section~\ref{s:agd}).}
If we assume a limiting line width of $\sim 1-2\rm\,km\,s^{-1}$, in the case of no turbulent
broadening, the corresponding CNM kinetic temperature is $\sim20-30\rm\,K$, which is also equal to the
spin temperature. The fact that 21-SPONGE and KOK14 
agree in their lower limit to $\Delta v_0$, together with the fact that the AGD method is a good 
measure of $T_{s,\rm true}$ at similar temperatures (c.f., Figure 4), suggest that the simulation and 
observations have a similar lower limit for the CNM temperature of $\sim 20-30\rm\,K$. 

As shown in Figure~\ref{f-cloud-hist}, the peak optical depth spans the whole parameter space
all the way to our sensitivity limit
with no obvious evidence for the existence of a minimum optical depth for the CNM.
In addition, only a small fraction, $<10\%$, of components have $\tau_0>1$ in KOK14 and 21-SPONGE.

\subsubsection{Role of the WF effect}
\label{abs-wf}

\begin{figure*}
	\vspace{-10pt}
   \includegraphics[width=\textwidth]{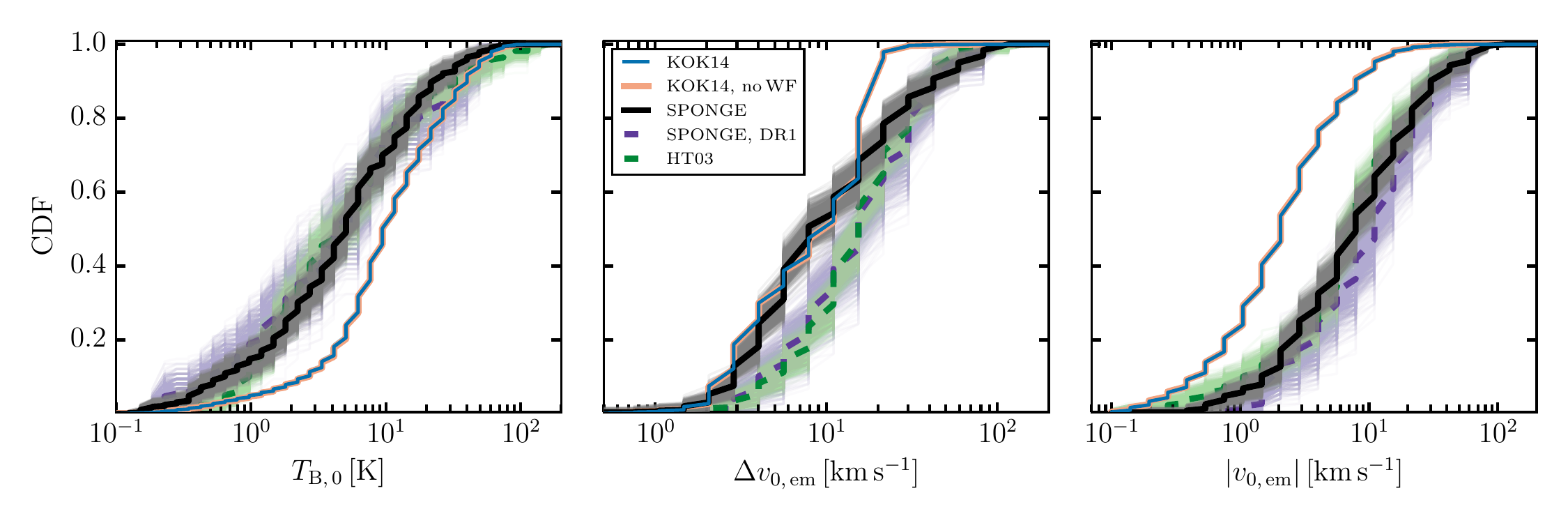}
   \caption{Cumulative distribution functions (CDFs) of Gaussian parameters, including brightness temperature amplitude ($T_{\rm B}$), FWHM ($\Delta v_0$) and mean velocity ($v_0$), of the components fitted to observed (21-SPONGE) and synthetic (KOK14) \hi\ emission spectra, including previous by-hand results from 21-SPONGE (purple, DR1; Murray et al. 2015) and HT03 (green) for comparison. For each dataset, we plot the CDFs of 1000 bootstrapped samples drawn from the full sample with replacement (shown in lighter-shaded colors according to the legend) to illustrate the effect of sample size and outliers on the CDF. }
      \label{f:gauss_em}
\end{figure*}

\begin{figure}
       \includegraphics[scale=0.55]{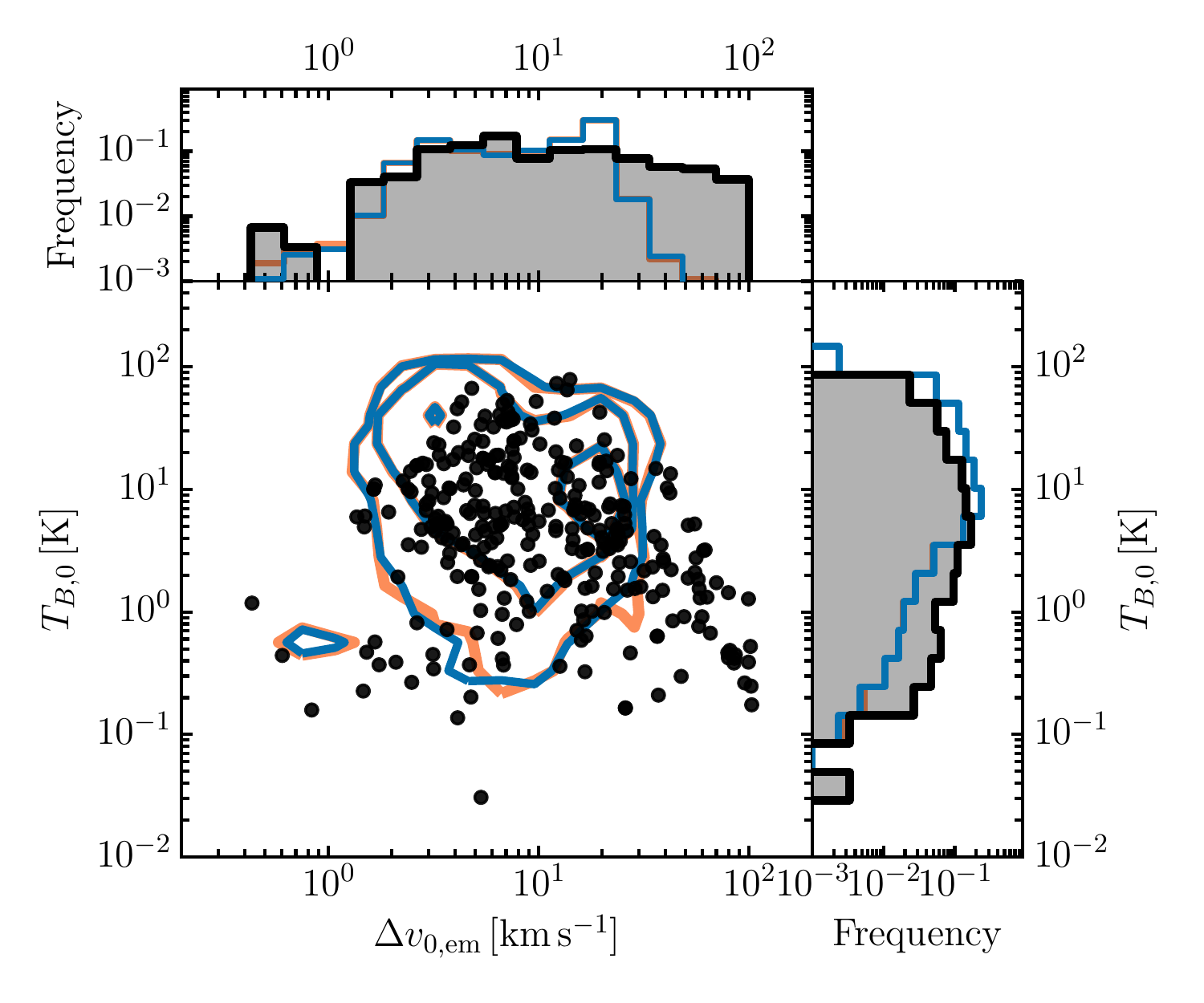}
   \caption{Parameters of Gaussian components (FWHM, $T_{B,0}$) fitted by AGD to \hi\ emission spectra from 21-SPONGE (black) and KOK14 with the WF effect (blue) and without the WF effect (orange). Contours indicate the 1, 2, and 3$\sigma$ limits for the KOK14 distributions. Marginal histograms display the same results according to the legend. }
      \label{f-cloud-em}
\end{figure}

As seen most clearly in the main panel of Figure~\ref{f-cloud-hist}, the KOK14 spectra with and without the WF effect include
significant populations of components with $\Delta v_0>10\rm\,km\,s^{-1}$ and $0.001<\tau_0<0.01$.
Although this region is located well above the 21-SPONGE median sensitivity in optical depth,
we find very few 21-SPONGE components there.
In addition, in KOK14, these components are often found without narrow 
(CNM-like) components superimposed along the same LOS. This is a type of profile not 
seen in 21-SPONGE observations. 
There are no observational biases that would prevent us from seeing simple Gaussian 
line profiles with a peak optical depth of $\sim0.01$ and a velocity FWHM of 10 km/sec.
In addition, the lack of isolated (devoid of CNM), broad, WNM-like features in observations is 
supported by additional high-sensitivity \hi\ absorption studies \citep[e.g., HT03, ][]{roy2013}.
An example isolated, broad absorption line from KOK14
is shown in the top row of Figure~\ref{f-cloud-matched-ex}. 

We note that these broad, low-$\tau$ features appear in the synthetic KOK14 
spectra regardless of our treatment of the WF effect, 
although agreement with observations is somewhat improved 
when this is included (c.f., Figure 11).  
The origin of these low-$\tau_0$ components is not well understood 
and future comparisons with synthetic $21\rm\, cm$ profiles from simulations will explore the effect of a 
more realistic feedback treatment \citep[e.g.,][]{kim2016sub}.
The present results already suggest that \hi\ absorption spectra may be able to 
provide discriminating tests of the input physics in simulations.

\subsection{Properties of \hi\ emission components}

Before comparing the properties of \hi\ emission components in the same
manner as in our comparison of absorption components in Section~\ref{abs-compare},
we emphasize again that the 21-SPONGE and KOK14 emission profiles are derived from
different angular scales. The angular resolution of the KOK14 emission spectra is the same as for the absorption lines.
However, the 21-SPONGE emission spectra were observed with the Arecibo radio telescope, with a $\sim3.5'$ beam at $21\rm \, cm$ derived from off-target positions, and therefore have different angular resolution than the 21-SPONGE VLA absorption spectra. 
In the case of the KOK14 spectra, which are derived from a simulation with
a physical resolution of $2\rm\,pc$, the path length would need to be longer
than $\sim2\rm\,kpc$ to achieve better angular resolution than 21-SPONGE. 
Therefore, for shorter LOS at predominately higher latitudes, the angular resolution of synthetic emission spectra
is actually worse than 21-SPONGE observations. 
We stress that absorption lines do not have the resolution problem and this also the reason why they are more suitable
for comparison with simulations.

Figure~\ref{f:gauss_em} displays CDFs of Gaussian
parameters from the fits to \hi\ emission spectra observed by 21-SPONGE (black), simulated by KOK14 
with the WF effect (blue) and without the WF effect (orange), in addition to the results 
of HT03 (green) and 21-SPONGE DR1 (purple).
These parameters include amplitude in brightness temperature ($T_{\rm B, \, 0}$ in $\rm K$; left), FWHM ($\Delta v_{0,\rm \, em}$ in $\rm km\,s^{-1}$; center) and mean velocity ($v_{0,\rm \, em}$ in $\rm km\,s^{-1}$; right). 
To illustrate the comparison between 21-SPONGE and KOK14 further,
we display $T_{B,0}$ vs. $\Delta v_{0,\rm em}$ for 21-SPONGE and KOK14
in Figure~\ref{f-cloud-em}, with marginal histograms for both parameters.
We note that the WF effect does not make a difference to components fitted to \hi\ emission spectra 
(i.e., orange and blue lines are indistinguishable in Figure~\ref{f:gauss_em}). 
This indicates that future testing of the implementation of the WF effect should use absorption, rather
than emission spectra.

\subsubsection{Boundaries in brightness temperature}

In the left panel of Figure~\ref{f:gauss_em} (as well as Figure~\ref{f-cloud-em}), the amplitudes fitted by 
AGD to 21-SPONGE agree well with 21-SPONGE DR1 and HT03. All three datasets were obtained using the Arecibo radio telescope,
and therefore they have similar angular resolution.
However, all three observed distributions are shifted to slightly lower amplitude in 
brightness temperature ($T_{B,0}<10\rm\,K$) relative to KOK14 in Figure~\ref{f:gauss_em}.
For the KOK14 LOS with lower effective angular resolution than 21-SPONGE, 
the synthetic brightness temperature spectrum averages any simulated emission over larger solid angles,
and therefore the KOK14 $T_{B,0}$ should tend to be smaller than the 21-SPONGE values 
derived from smaller angular scales. However, we observe more low-$T_{B,0}$
components in 21-SPONGE than KOK14.

The slight excess of components with high $T_{B,0}$ and low $\Delta v_0$ in KOK14 
may be caused by the lack of chemistry and \hi-to-H$_2$ transition in the simulation.  
Furthermore, as noted previously, the relative lack of high-$\Delta v_0$ components in 
KOK14 may be partly attributed to the reduced velocity dispersion in the simulations compared to observations.
Our observations may even suggest an upper-limit on $T_{B,0}$ of $\sim 60-70$ K.
In addition, on the low-end of the distribution
the absence of a large filling-factor ($\sim 50\%$) hot ($T\sim10^5-10^7\rm\,K$) medium in the simulations
produces too much neutral gas per LOS, and therefore too few LOS with $T_{B,0}<$ a few K. 
The presence of a hot medium occupying a large fraction of the volume would also tend to 
reduce the incidence of detectable low-$\tau_0$ features, since many of the LOS without 
CNM would be primarily hot rather than primarily warm medium.

\subsubsection{Broad emission components}

In the middle panel of Figure~\ref{f:gauss_em}, the 21-SPONGE and KOK14 $\Delta v_{0,\rm em}$ 
distributions agree very well below $\Delta v_{0,\rm em}\sim 10\rm\,km\,s^{-1}$. However, the 21-SPONGE AGD, DR1 and HT03 distributions
extend to higher values of $\Delta v_{0,\rm em}$ than KOK14.
This is especially noticeable in Figure~\ref{f-cloud-em} where 21-SPONGE components 
with large $\Delta v_{0,\rm em}$ and small $T_{B,0}$ form a prominent
tail of the distribution. As discussed previously, stray radiation
the 21-SPONGE \hi\ emission observations have not been corrected for stray radiation.
which would appear in the form of weak and broad spectral components. To test how many
21-SPONGE components could be affected by stray radiation, we extracted \hi\ emission spectra
from the stray-radiation-corrected LAB survey \citep{kalberla2005} at the positions of our sources and 
implemented AGD in same manner to decompose the LAB spectra into Gaussian components.
Consequently, LAB data contain many components with $\Delta v_{0,\rm em}\sim10-70\rm\,km\,s^{-1}$ (c.f., Appendix).
However, components with $\Delta v_{0,\rm em}\gtrsim70\rm\,km\,s^{-1}$ are not seen in LAB data, and therefore
those components in 21-SPONGE are likely caused by stray radiation. 
There are 9 such components and they are all located in the
large $\Delta v_{0,\rm em}$ and small $T_{B,0}$ tail. \citet{peek2011},
did a similar comparison between LAB and Arecibo data and concluded that the effect is unlikely
to exceed $500\rm\, mK$, in agreement with our with our conclusion that the broadest 21-SPONGE components with $T_{B,0}<0.5$ K are likely caused by stray radiation.

This leaves a population of broad ($\Delta v_{0,\rm em}\gtrsim20-70\rm\,km\,s^{-1}$), shallow ($T_{B,0}<10\rm\,K$) 
components that are prominent in 21-SPONGE but absent in KOK14.
Possibly a higher velocity dispersion and the inclusion of Galactic fountain in the simulation \citep[see e.g.,][]{kim2016sub} may be able
to reproduce such broad lines. Alternatively, this could signify the presence of the WNM at temperature
$>4000$ K. For example, \citet{murray2014} detected a residual absorption
component with a width of 50 km/s and $T_s\sim7000\rm\, K$ by stacking the first third of the 21-SPONGE data. 
Further work is clearly needed to understand the origin of 
such broad emission components, as well as to remove stray radiation from Arecibo spectra.

Finally, in the right panel of Figure~\ref{f:gauss_em}, all observed \hi\ emission spectra 
have absolute mean velocities up to $\sim 50\rm\,km\,s^{-1}$, whereas 
the KOK14 mean velocities appear limited to $v_{0,\rm em}<10\rm\,km\,s^{-1}$ 
(c.f., right panel Figure~\ref{f:gauss_abs}).
In addition to the nature of the local box reducing $|v_{0,\rm em}|$, the KOK13 
simulations did not include a Galactic fountain of WNM with velocities up to 
tens of $\rm km\,s^{-1}$. The inclusion of this important mechanism may 
produce relatively more structures with larger $|v_{0,\rm em}|$ and $\Delta v_{0,\rm em}$ 
and improve similarities between the 21-SPONGE and simulation results. 

\begin{figure}
   \centering
   \vspace{10pt}
   \includegraphics[scale=0.7]{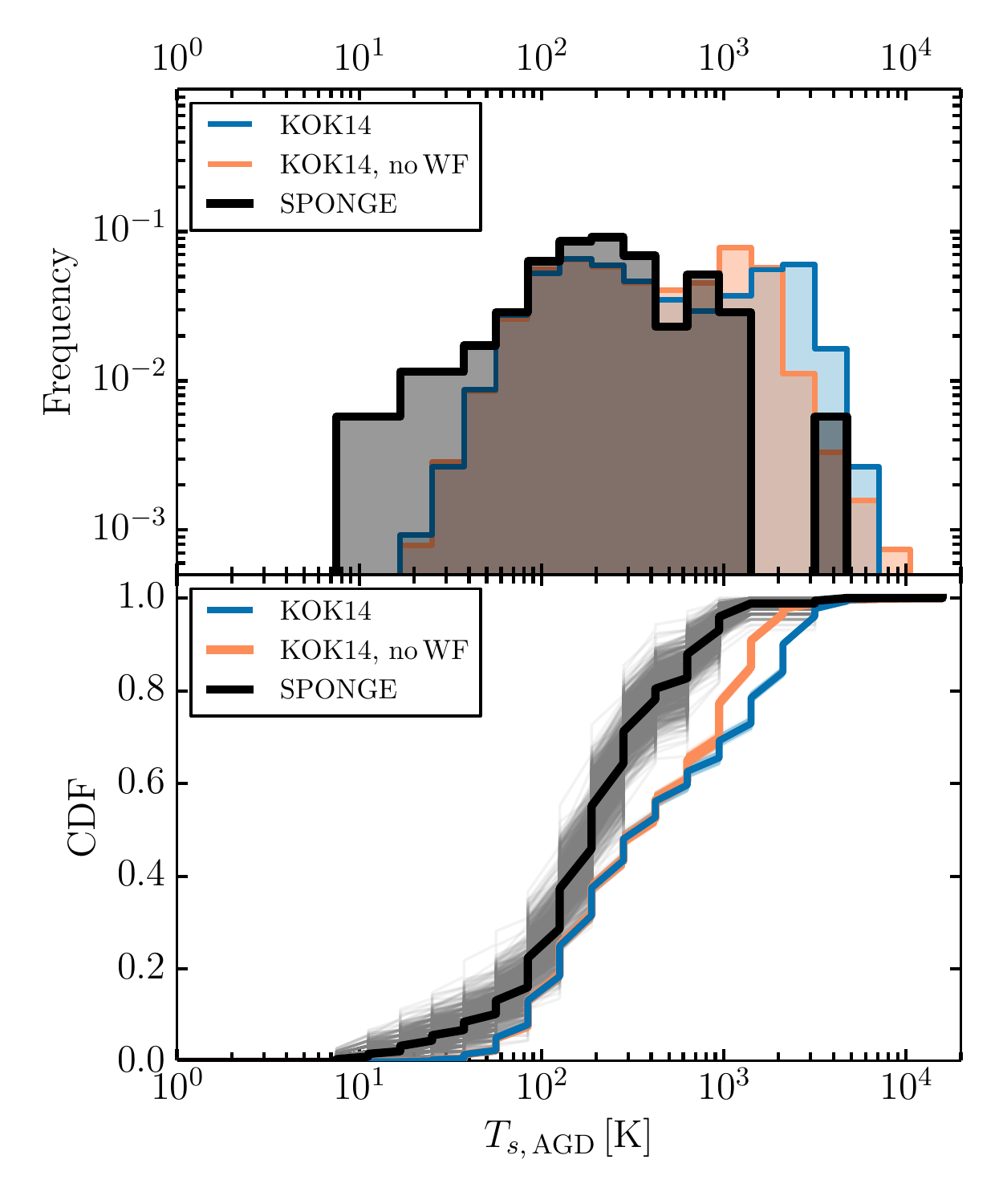}
   \caption{
   Top: histograms of $T_{s,\,\rm AGD}$ (Equation~\ref{ts-agd}) for all components which ``match" between \hi\ emission and absorption (Equations~\ref{sigs_away} and \ref{fwhm_factor}) following the AGD analysis of  \numbersponge{} 21-SPONGE \hi\ spectral pairs (thick, solid black) and \numberkok{} KOK14 \hi\ spectral pairs with the WF effect (thin, solid blue) and without the WF effect (thin, solid orange). Bottom: CDFs of the same results, with 1000 bootstrapped samples drawn from the full sample with replacement (shown in lighter-shaded colors according to the legend) to illustrate the effect of sample size and outliers on the CDF.    }
   \label{ts-histo}
   \vspace{10pt}
\end{figure}

\subsection{Observed Spin Temperature}
\label{s-obsim-temp}

We now compare distribution functions of the inferred spin temperature,
from 21-SPONGE spectra and KOK14 synthetic spectra. 
As we have shown in Section~\ref{s-spintemp}, for the majority of cases the inferred spin temperature
using our AGD and radiative transfer approach, $T_{s,\rm AGD}$,  is in agreement with the
true simulated temperature, $T_{s,\rm true}$. We also discussed how at $>400\rm\,K$, $T_{s,\rm AGD}$ 
over-estimates $T_{s,\rm true}$. However, this bias will affect both 
observations and simulations in the same way, as we apply the same 
AGD fit and radiative transfer method to KOK14 and 21-SPONGE.
In addition, our main focus in this paper is on the shape of distribution functions, not the exact 
fractions. In future work, we will focus on the fractions of \hi\ in CNM, WNM and unstable phases using updated simulations.

Figure~\ref{ts-histo} displays histograms and CDFs of 
$T_{s,\rm AGD}$ (Equation~\ref{ts-agd}) for 21-SPONGE (black) and KOK14 with (blue) and without (orange) the WF effect.
The observed and simulated $T_{s,\rm AGD}$ distributions follow each other well
until $T_{s,\rm AGD}\sim400-500\rm\, K$, when they start to diverge. 
Although the 21-SPONGE observations appear to have a higher relative fraction of 
components at low $T_{s,\rm\,AGD}\sim20-30\rm\,K$, these bins are determined by small
number of components, and the fractions agree within uncertainties as illustrated
by the bootstrapped samples shown in the CDF panel. 
As discussed in Section 6.2, the observations and simulations display
consistent cutoff in CNM line width, and Figure~\ref{ts-histo} indicates
that the fractions of material at corresponding spin temperatures of $20-30\rm\,K$
are consistent. HT03 detected a similar population of cold CNM components with $\lesssim20\rm\,K$ ($17\%$ by number,
and $4\%$ by mass), and suggested that this is evidence for the absence of 
photoelectric heating by dust \citep{wolfire1995}.
We note that the simulations have uniform heating throughout and constant metallicity, 
while in reality the photoelectric heating would be reduced in high-column regions.  

\begin{figure}
   \centering
   \vspace{10pt}
   \includegraphics[scale=0.6]{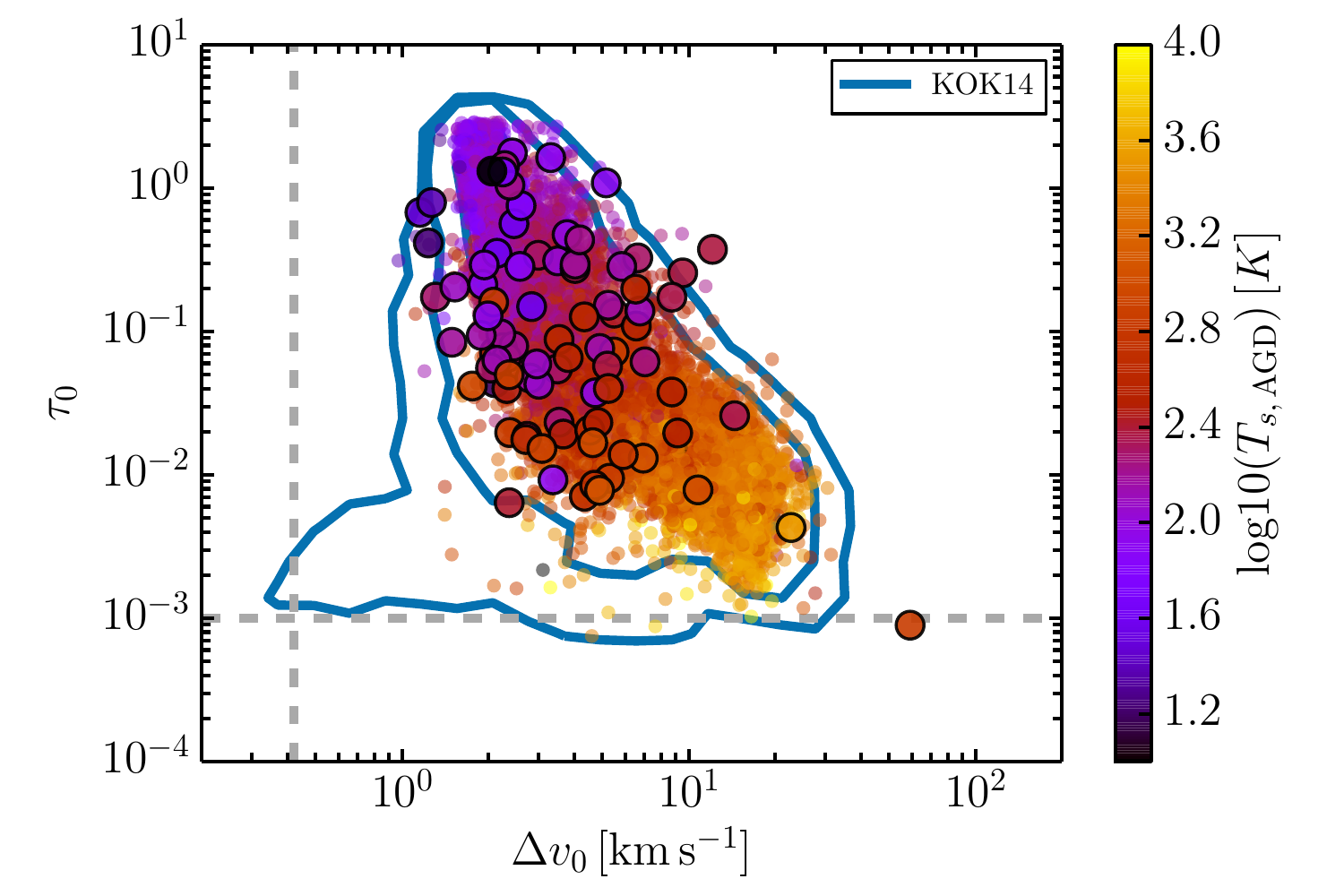}
   \caption{AGD absorption properties ($\tau_0$, $\Delta v_0$) for all components fitted to KOK14 with the WF effect (blue contours), and all components which match between absorption and emission according to Equations~\ref{sigs_away} and~\ref{fwhm_factor} for KOK14 (small circles) and 21-SPONGE (large circles with black outlines), colored by the AGD spin temperature ($T_{s,\rm AGD}$). 
}
   \label{bluecloud_match}
   \vspace{10pt}
\end{figure}

However, the KOK14 spectra show more \hi\ with $T_{s,\rm AGD}=300-3000$ K
than 21-SPONGE. To illustrate the types of components with these temperatures,
in Figure~\ref{bluecloud_match} we reproduce Figure 11, including the total contours with WF from KOK14 
and all matched components from KOK14 (small circles0 and 21-SPONGE (large circles with black outlines), colored
by $T_{s,\rm AGD}$. The AGD components with $T_{s,\rm AGD}\sim300-3000\rm \, K$ in the KOK14 have lowest optical depth and the largest FWHM. These are exactly the components discussed in Section~\ref{abs-compare} to have simple, broad profiles without overlapping CNM components -- a type of spectral profile which is not found often in 21-SPONGE.

\begin{figure}
   \centering
   \vspace{10pt}
   \includegraphics[scale=0.7]{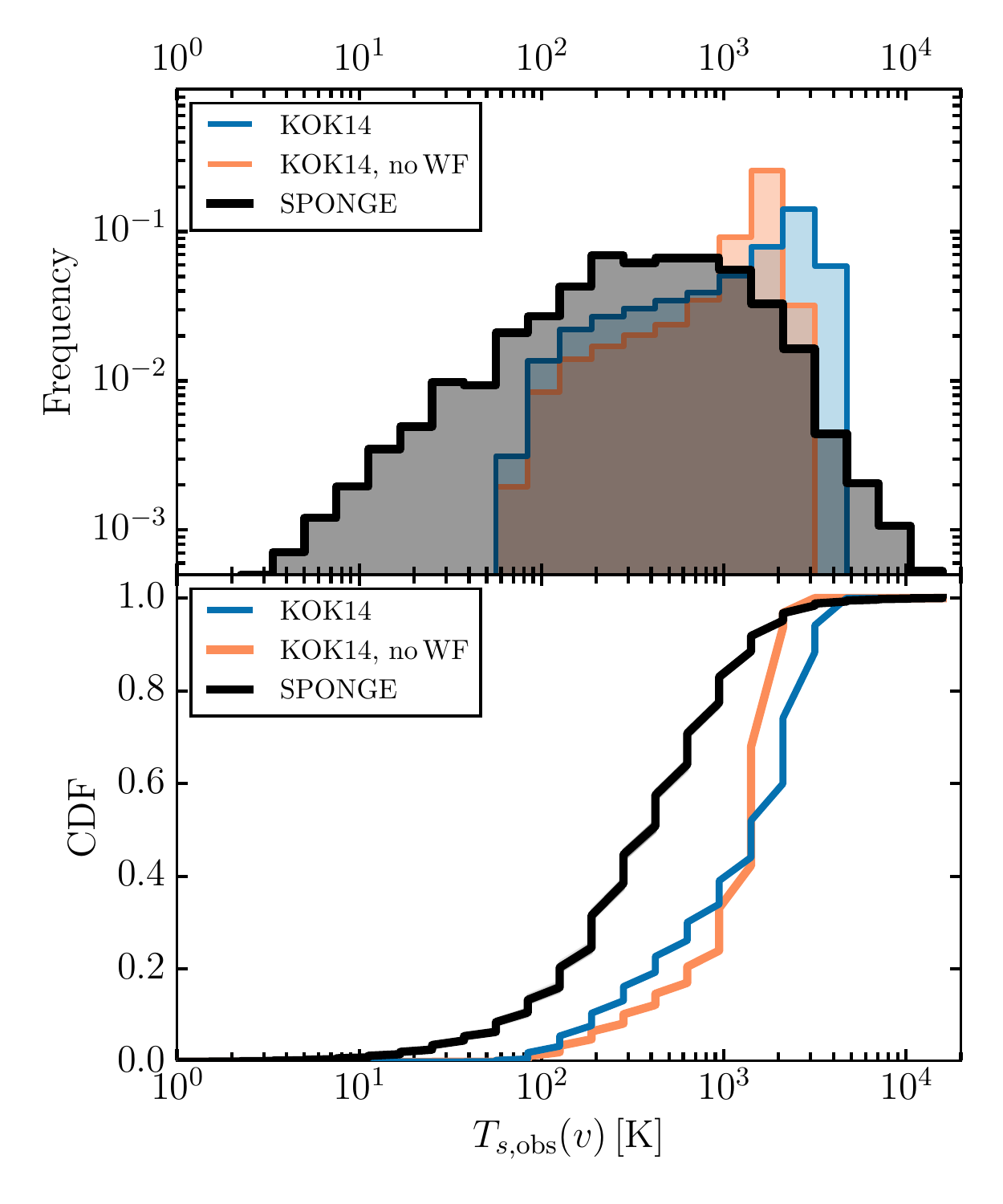}
   \caption{
   Top: histograms of per-channel spin temperature ($T_s(v)$) for all \numbersponge{} 21-SPONGE \hi\ spectral pairs (thick, solid black) and \numberkok{} KOK14 \hi\ spectral pairs with the WF effect (thin, solid blue) and without the WF effect (thin, solid orange).
    Bottom: CDFs of the same results.  }
   \label{ts-histo-perch}
   \vspace{10pt}
\end{figure}

\begin{figure}[t!]
   \centering
   \vspace{10pt}
   \includegraphics[scale=0.7]{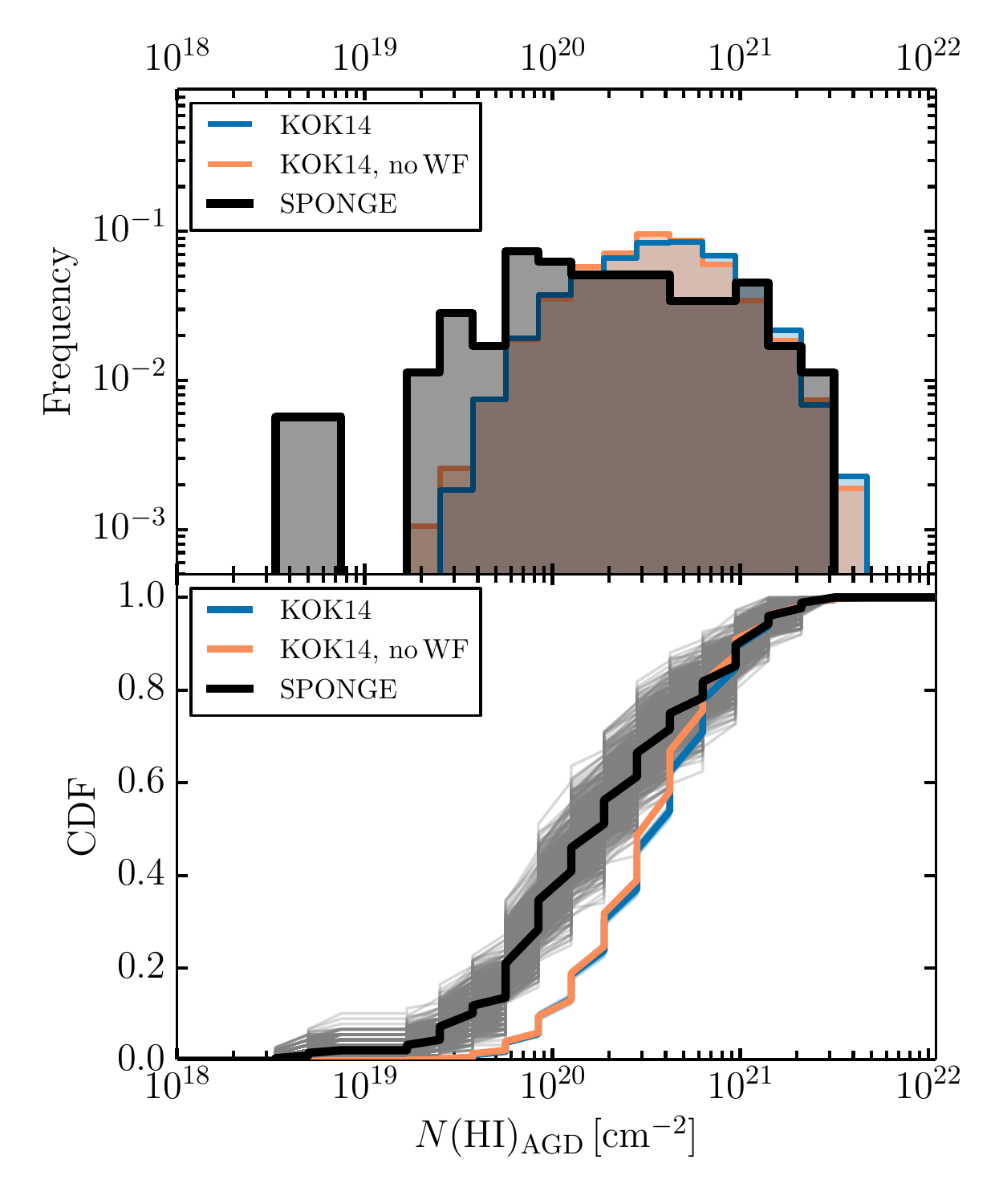}
      \caption{Top: histograms of $N({\rm H\textsc{i}})_{\rm AGD}$ (Equation~\ref{nhcomp}) for all components which ``match" between \hi\ emission and absorption (Equations~\ref{sigs_away} and \ref{fwhm_factor}) following the AGD analysis of  \numbersponge{} 21-SPONGE \hi\ spectral pairs (thick, solid black) and \numberkok{} KOK14 \hi\ spectral pairs with the WF effect (thin, solid blue) and without the WF effect (thin, solid orange). 
Bottom: CDFs of the same results, with 1000 bootstrapped samples drawn from the full sample with replacement (shown in lighter-shaded colors according to the legend) to illustrate the effect of sample size and outliers on the CDF. 
}
      \vspace{10pt}
   \label{nh-histo}
\end{figure}

Are the differences shown in Figures~\ref{ts-histo} and~\ref{bluecloud_match} caused by observational sensitivity? 
21-SPONGE was designed with goal of detecting warm \hi\ in absorption and has excellent optical depth sensitivity \citep[RMS noise $\sigma_{\tau}<10^{-3}$ per channel;][]{murray2015}. 
To estimate the temperatures we are sensitive to in 21-SPONGE observations, 
we assume a WNM column density of a few $\times 10^{20}$ \citep[e.g., ][, KOK14]{stanimirovic2014},
a FWHM of $10-20\rm\, km\,s^{-1}$ and a conservative RMS sensitivity in optical depth
of $10^{-2} - 3\times 10^{-3}$ (per 0.4 km/s velocity channels), which results in $T_s\sim1000-6000\rm\,K$. 
This is the range of what is expected by KOK14 and also \citet{liszt2001}.
Therefore, 21-SPONGE has the observational sensitivity to detect spectral components with $T_s>1000\rm\,K$.
In Section~\ref{s-spintemp}, we demonstrated that our AGD and radiative transfer implementation
are not biased against WNM, and in fact above $\sim400\rm\,K$ tends to over-estimate the true spin temperature.
Therefore, we infer that the lack of observed components with $T_s>1000$ K is 
not affected by sensitivity or analysis method. In addition, we emphasize again that the $T_s>1000$ K components in 
simulations have isolated, simple spectral profiles that we do not find in observations.

To further test the effect of observational sensitivity on the inferred
spin temperature, \citet{murray2014} investigated methods for improving the 21-SPONGE
sensitivity to broad, shallow WNM-like absorption with high temperature. We stacked residual \hi\ absorption 
lines of a subset of sources following by-hand Gaussian decomposition and
found a residual \hi\ absorption signal at $5\sigma$ significance with an inferred excitation
temperature of $T_s=7200_{-1200}^{+1800}\rm\,K$,
with a FWHM of $50\rm\,km\,s^{-1}$, and an \hi\ column density of $2\times 10^{20}$ cm$^{-2}$ \citep{murray2014}.
This temperature is higher than analytical predictions for collisional excitation of \hi, 
and indicated that additional \hi\ excitation mechanisms (e.g., the WF effect) 
may be more important for coupling the hyperfine spin
states of \hi\ to the local thermodynamic temperature than previously thought.

If the spin temperature of the WNM is actually $T_s\sim7000\rm\,K$ \citep{murray2014},
rather than $2000<T_s<4000\rm\,K$ as inferred currently from analytical models 
of collisional excitation \citep[e.g.,][]{liszt2001},
then it is possible that individual 21-SPONGE spectra may still lack sensitivity for detecting the WNM.
Even more sensitive observations, as well as further stacking analysis will be essential from 
an observational standpoint. This result shows that understanding the implementation of the WF effect 
in numerical simulations, and in particular
the highly uncertain $n_{\alpha}$, its spatial variations across the Milky Way, as well
as the effect of turbulence, is essential to reconcile \hi\ observations and theory.

\subsubsection{Per-channel $T_s$}

Instead of using Gaussian-based temperature estimates, 
KOK14 considered per-channel spin temperature in their analysis of simulated \hi\ properties, 
They found that this quantity agrees well with the true per-channel temperature extremely well (within a factor of 1.5) for
all channels with $\tau\lesssim 1$ (KOK14). 
Several observational studies have also used per-channel temperature estimates to analyze \hi\ phases \citep[e.g.,][]{roy2013}.
To compare with their results, we derive per-channel spin temperature, $T_{s,\rm obs}(v)$, by applying an equation similar to Equation~\ref{ts-perch}
to the full $\tau(v)$ and $T_B(v)$ spectra from 21-SPONGE and KOK14, where,
\begin{equation}
T_{s,\rm obs}(v) = \frac{T_B (v) }{1- e^{-\tau(v)}}.
\label{ts-obs-perch}
\end{equation}

\noindent For each LOS, we compute $T_{s,\rm obs}(v)$ for only those channels with
optical depths greater than $3\times 10^{-3}$, to conservatively exclude all channels with 
low S/N. 

In Figure~\ref{ts-histo-perch}, we display histograms and CDFs of $T_{s,\rm obs}(v)$ for 21-SPONGE (black)
and KOK14 with (blue) and without (orange) the WF effect. 
There is a similar discrepancy between observations and simulations in Figure~\ref{ts-histo-perch} as seen in Figure~\ref{ts-histo}. 
The KOK14 distributions are shifted to higher temperatures, while 21-SPONGE
spectra contain more channels with low temperature ($<200\rm\,K$). In addition, the 21-SPONGE CDF has a more gradual 
rise, while simulated data show an abrupt jump near $T_{s,\rm obs}(v)\sim1000\rm\,K$ suggesting
that most simulated LOSs are dominated by higher $T_{s,\rm obs}(v)$ derived from spectral channels without detectable absorption.
It is important to keep in mind, however, that observed and simulated spectra probe different LOS lengths, as
discussed in Section~\ref{abs-compare}, which could affect the shape of per-channel CDFs. 
In addition, the absence of a hot medium in the present simulations may lead to too many LOS with $T_s \sim 1000\rm\,K$.  

\subsection{Observed Column Density}

As a further benefit of AGD analysis, by resolving the properties of individual spectral components 
along each LOS, we can analyze the column densities of individual gas structures 
in contrast with the total LOS column density.  
In Figure~\ref{nh-histo}, we display histograms and CDFs of $N({\rm H\textsc{i}})_{\rm AGD}$ 
for individual matched spectral components from the KOK14 (blue) and 
21-SPONGE (thick black) \hi\ emission and absorption spectral pairs. The column 
density distributions shown in Figure~\ref{nh-histo} agree well at 
high-$N(\rm H\textsc{i})$ ($>10^{20}\rm\,cm^{-2}$), 
 although the 21-SPONGE distribution extends further below $N(\rm H\textsc{i})=10^{19}\rm\,cm^{-2}$. 
 
The absence of low-column lines in the KOK14 spectra may be caused by insufficient angular resolution for 
detecting small CNM features, or, as has been mentioned throughout, the absence of a hot medium in the KOK13 simulations
which would serve to reduce the observed column densities of the matched lines.
The discrepancy around $10^{20}$ cm$^{-2}$ is likely caused by the discrepancy in $T_{s,\rm AGD}$ 
discussed above in Section~\ref{s-obsim-temp}.
It is interesting to note that the application of the WF effect does
not significantly affect the $N(\rm H\textsc{i})_{\rm AGD}$ distribution.
The WF effect influences the optical depth and spin temperature in opposite ways, i.e., increases $T_s$ and decreases $\tau$. These quantities are the main ingredients of $N(\rm H\textsc{i})_{\rm AGD}$ (c.f., Equation~\ref{nhcomp}), and
therefore the two results of the WF effect may cancel each other out when computing component-based column density.
We conclude that the complexity of factors incorporated into the \hi\ column density make it a less useful tool for isolating the importance of the WF effect.

\section{Summary and Conclusions}
\label{s:conclusions}

Detailed comparisons between observations and simulations are crucial for understanding the physics
behind the observed properties of the ISM. Armed with synthetic $21\rm\,cm$ emission and absorption profile data created from the 3D hydrodynamical simulations from KOK14 and 
high-sensitivity \hi\ observations from 21-SPONGE, we address two main questions:
(1) how well do \hi\ spectral lines and our analysis methods recover simulated properties of interstellar gas structures? (2) how do simulated \hi\ spectra compare with real observations?
To analyze \numberkok{} synthetic and \numbersponge{} real observations in an 
unbiased and uniform way, we apply the Autonomous Gaussian Decomposition (AGD) algorithm \citep{lindner2015} 
identically to both datasets. With these fits in hand, we compare simulated properties of gas structures along each LOS with 
observed properties of the Gaussian components. 

We summarize the main results:

\begin{enumerate}

\item For gas structures defined by peaks in $n/T_s$ along random LOS in the KOK13 simulations, Gaussian fits by AGD to synthetic \hi\ absorption lines are able to recover gas structures successfully (Figure~\ref{f-cloud-matched}). The recovery completeness (Equation~\ref{e:complete}) is $99\%$ for high-latitude LOS ($|b|>50^{\circ}$), $67\%$ for mid-latitude LOS ($20<|b|<50^{\circ}$) and $53\%$ for low-latitude LOS ($0<|b|<20^{\circ}$). The completeness declines with decreasing latitude because the LOS complexity is highest at the lowest latitudes. When these structures are matched to spectral line components in both \hi\ absorption and emission, the completeness is $83\%$, $38\%$ and $29\%$ for high, mid and low latitudes respectively. The decline in recovery completeness when matches between gas structures and both \hi\ absorption and emission components are required reflects the difficulty in associating unambiguous spectral features in the presence of line blending and turbulence. 

\item We use AGD fits to synthetic lines and simple radiative transfer to compute observational estimates of spin temperature ($T_{s,\rm AGD}$) and column density ($N({\rm H\textsc{i}})_{\rm AGD}$) for matched pairs of \hi\ absorption and emission lines. We compare these estimates with the simulated spin temperatures ($T_{s,\rm true}$) and column densities ($N({\rm H\textsc{i}})_{\rm true}$) of corresponding structures in the simulation. The observed and simulated spin temperatures agree within a factor of 2 for the majority of structures ($68\%$; Figure~\ref{f-temp-compare}). At high temperatures, $T_{s,\rm AGD}$ overestimates the $T_{s,\rm true}$ due to velocity offsets between \hi\ absorption and emission lines caused by turbulent motions (Figure~\ref{f-inspect}).

The observed and simulated \hi\ column densities also agree well for the majority of structures. However, the scatter is slightly larger than in the case of spin temperature, because $N(\rm H\textsc{i})_{\rm AGD}$ incorporates all uncertainty in $T_{s,\rm AGD}$ (Figure~\ref{f-nh-compare}). Furthermore, the agreement between inferred and true properties declines at low Galactic latitude and for low-$\tau$ components, where LOS-blending components hinder clear associations between emission and absorption spectral lines (Figure~\ref{f-analyze}). 

Overall, the agreement between temperature and column densities inferred from
synthetic spectra and computed from physical conditions in the simulation is encouraging. Future comparisons with next-generation simulations 
will allow us to construct ``correction functions" for observed spin temperature and column density 
distributions. 

\item We find more fitted absorption and emission lines per LOS ($N_{\rm AGD}$) in the 21-SPONGE observations than the KOK14 synthetic observations (Table~\ref{t:AGD}). This difference reflects the fact that the simulated scale heights of the CNM and WNM in the KOK13 simulations are lower than in observations, due to velocity dispersions lower than seen in observations ($\sim 5-7\rm\,km\,s^{-1}$). These results are derived from identical implementation of AGD to 21-SPONGE and KOK14, and thus the comparison is unaffected by biases introduced by AGD analysis. The discrepancy reflects the limitations of local box simulations, with a simplified treatment of supernova feedback, in producing realistic synthetic spectral lines. 

In addition, there are comparatively fewer matches per LOS in 21-SPONGE than KOK14. The so-called mismatch in angular resolution between 21-SPONGE \hi\ emission ($\sim3.5'$) and absorption ($1-40''$) complicates the matching process described by Equations~\ref{sigs_away} and~\ref{fwhm_factor}. In the future, we plan to quantify this effect and correct the observational results by smoothing simulated spectra.

\item Using AGD, we objectively compare the properties of spectral lines fitted by AGD to \hi\ absorption from 21-SPONGE and KOK14. The 21-SPONGE spectra have a wider range in mean velocity (right panel, Figure~\ref{f:gauss_abs}), due to the limited horizontal box size of the simulation. Furthermore, at high Galactic latitudes where the influence of the global effect is weakest, $N_{\rm AGD}$ and matching statistics agree between 21-SPONGE and KOK14. This indicates that simulated Galactic rotation plays an important role in observed \hi\ properties, and improved implementations of global effects will improve the completeness statistics of \hi\ structure recovery, and improve the completeness characterization discussed in Summary points 1, 2 and 3. 

\item We find that KOK14 spectra include more low-$\tau$ and high-$\Delta v$ absorption lines than are seen in 21-SPONGE (Figures~\ref{f:gauss_abs} and~\ref{f-cloud-hist}), despite being well above the 21-SPONGE sensitivity and resolution limits. These broad spectral lines are often found without narrower, blended lines, which is a profile not seen in observations by 21-SPONGE or previous \hi\ absorption line surveys (e.g., HT03, Roy et al.\,2013). These are likely the result of the absence of a hot, large filling-factor gas phase in the KOK13 simulations (which would increase the number of LOS without detectably-absorbing neutral gas), or possibly some aspect of the simple treatment of the WF effect. In particular, we find that excluding the WF effect enhances the population of these discrepant \hi\ absorption components, and suggests that the WF effect is important for realistic spectral line properties. These features are not obvious in comparisons of integrated or per-channel properties, and reflect the utility of studying velocity-resolved spectral components. 

\item We find that \hi\ absorption spectra are more useful probes of ISM physics in comparison with simulations than \hi\ emission spectra. Properties of components fitted to \hi\ emission profiles are affected by angular resolution mismatch and stray radiation, and are not sensitive to the implementation of the WF effect. 

\item The AGD-derived spin temperature from KOK14 has more high-temperature gas ($1000<T_{s,\rm AGD}<4000\rm\,K$) relative to 21-SPONGE. The AGD method is not biased against high temperatures, and the 21-SPONGE observations have sufficient sensitivity for detecting gas at similar values. In KOK14, the highest-temperature components have large line widths and low optical depths, which is a type of profile not seen in 21-SPONGE (Figure~\ref{bluecloud_match}). We conclude that the lack of observed components with $T_s > 1000\rm\, K$ is not affected by sensitivity or the analysis method. We suggest that this gas may have even higher temperatures ($T_s\sim7000\rm\,K$) than what 21-SPONGE is sensitive to and than what the KOK14 implementation of the WF effect allows (i.e., $1000<T_s<4000\rm\,K$). Again, future work is required to understand the importance of the WF effect in more detail. It is also important to test whether more realistic treatment of star formation feedback, including a hot ISM \citep[see][]{kim2016sub}, reduces the incidence of features with $T_{s,\rm AGD} \sim 1000-4000\rm\, K$ in simulations.  

\end{enumerate}

Overall, we are encouraged that the AGD analysis and radiative transfer method presented here 
is a useful tool for diagnosing important physical 
conditions within ISM simulations. 
In the future, we will apply the strategies described here to updated
simulations in order to derive correction functions for observational biases in real \hi\ data. 
Upcoming large \hi\ absorption surveys such as GASKAP at the Australian Square
Kilometer Array Pathfinder (ASKAP) telescope \citep{dickey2013} will contribute many more sources
and improve the observational statistics. The objective and efficient nature of the AGD analysis strategy presented here is  
well-suited for future large observed and simulated datasets, and will be important for 
understanding the balance of CNM and WNM in the local and extragalactic ISM. 

\acknowledgements
This work was supported by the NSF Early Career 
Development (CAREER) Award AST-1056780. C.~E.~M. acknowledges 
support by the National Science Foundation Graduate Research 
Fellowship and the Wisconsin Space Grant Institution. 
S.~S. thanks the Research Corporation for Science 
Advancement for their support.  
The work of E.C.O. and C.-G. Kim was supported by NSF grant AST-1312006.
The National Radio Astronomy Observatory is a facility of the National Science 
Foundation operated under cooperative agreement by Associated 
Universities, Inc. The Arecibo Observatory is operated by 
SRI International under a cooperative agreement with the 
National Science Foundation (AST-1100968), and in alliance 
with Ana G. M\'endez-Universidad Metropolitana, and the 
Universities Space Research Association. 

\bibliographystyle{apj}
\bibliography{ms}

\appendix{}

\begin{figure*}[b!]
       \includegraphics[scale=0.55]{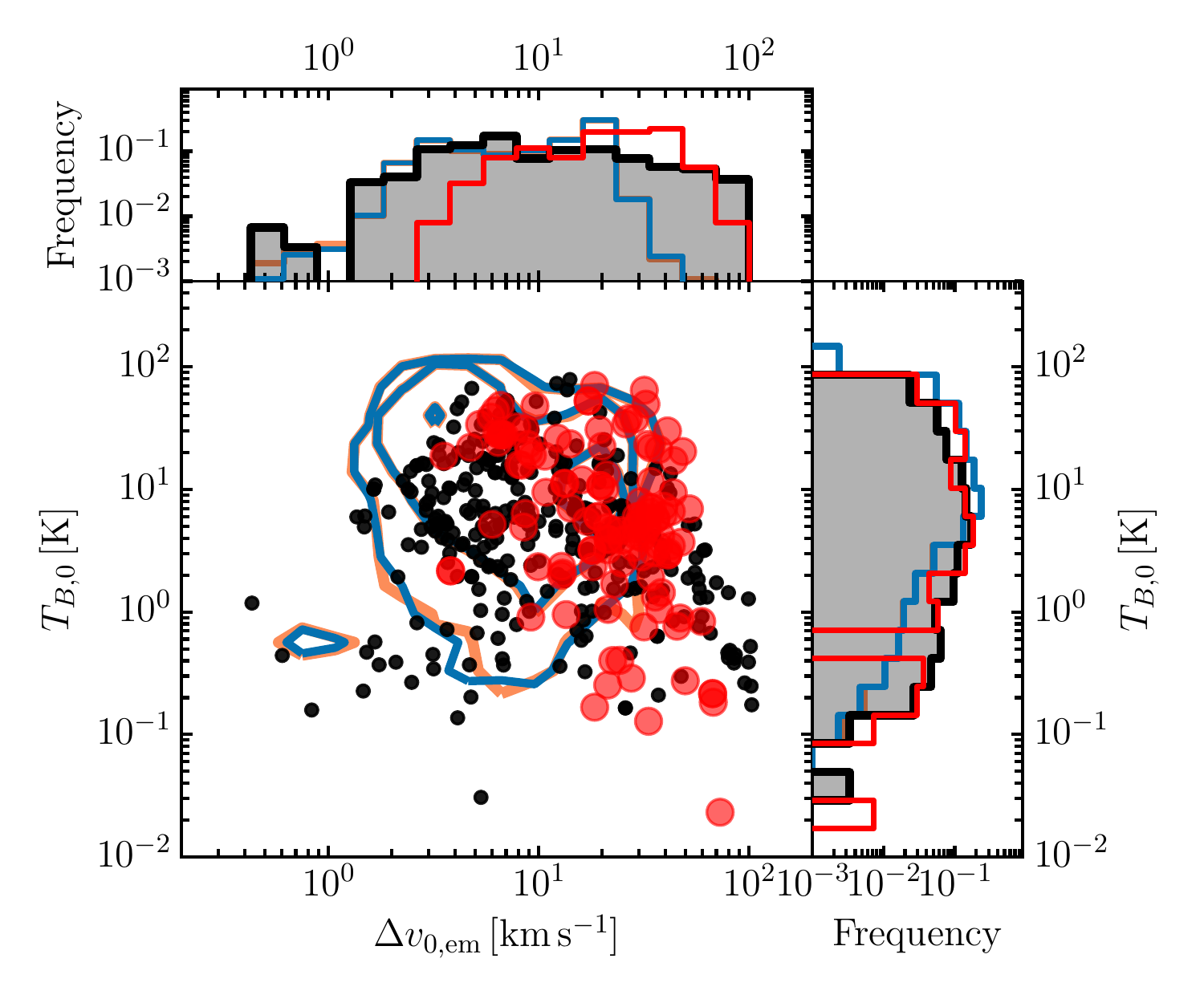}
   \caption{Parameters of Gaussian components ($\Delta v_{0,\rm em}$, $T_{B,0}$) fitted by AGD to \hi\ emission spectra from 21-SPONGE (black) and KOK14 synthetic observations including the WF effect (blue) and without the WF effect (orange). Contours indicate the 1, 2, and 3$\sigma$ limits for the KOK14 distribution. We also include the AGD decomposition of \hi\ emission spectra in the direction of the 21-SPONGE sources from the Leiden Argentine Bonn (LAB) survey \citep{kalberla2005} in red. }
      \label{f-cloud-em-lab}
\end{figure*}

To test for the effect of stray radiation 
on \hi\ emission spectra, we analyze data from the LAB survey \citep{kalberla2005}. The LAB spectra were measured with a $\sim 35'$ beam, whose shape was carefully modeled in order to remove contamination from stray radiation. We apply the AGD algorithm in the same manner as described in Section 6.3 to LAB spectra in the direction of the \numbersponge{} 21-SPONGE sources. In Figure~\ref{f-cloud-em-lab}, we reproduce Figure~\ref{f-cloud-em} including the LAB decomposition in red. 

From Figure~\ref{f-cloud-em-lab}, the LAB $\Delta v_{0,\rm em}$ distribution is shifted to larger values relative to KOK14 and 21-SPONGE. As a result of the much larger angular resolution of the LAB survey, the AGD decomposition of LAB data in Figure~\ref{f-cloud-em-lab} feature fewer narrow velocity components ($\Delta v_{0,\rm em}<2\rm\,km\,s^{-1}$), likely because the \hi\ emission is smoothed over larger angular scales than in the 21-SPONGE spectra (from Arecibo Observatory, with $\sim3.5'$ angular resolution). 

Furthermore, the LAB decomposition contains large-line width ($\Delta v_{0, \rm em}>30\rm\,km\,s^{-1}$), low-brightness temperature ($T_{B,0}<1\rm\,K$) components similar to 21-SPONGE. However, there is a population of 21-SPONGE components with $\Delta v_{0,\rm em}\sim 50-100\rm\,km\,s^{-1}$ outside the LAB distribution. Given that stray radiation has been removed from the LAB spectra, this type of spectral feature may be indicative of stray radiation. 

Future work is needed to understand and remove the stray radiation contamination from Arecibo spectra. 
 \citet{kalberla2010} found that stray radiation contributed up to $35\%$ to some GASS \hi\ spectra, however
the effect is typically $<10\%$ and is not significant at Galactic latitudes below $\sim60^{\circ}$ \citep[e.g.,][]{mcg2009}. 
In \citet{murray2015}, we compared 21-SPONGE emission spectra with the GALFA-\hi\ \citep{peek2011} and LAB surveys \citep{kalberla2005} and concluded that difference are generally within $3\sigma$ uncertainties.
In a future paper, we will compare the KOK14 synthetic \hi\ emission spectra and data from the LAB and GALFA-HI surveys objectively using AGD in order to statistically quantify the effect of stray radiation on observational data. 

\end{document}